\documentclass{jfm}
\usepackage{graphicx}
\usepackage{bm}
\usepackage{epstopdf, epsfig}
\usepackage{ upgreek }
\usepackage{csquotes}
\usepackage{xcolor}
\usepackage[normalem]{ulem}
\usepackage{url}

\newcommand{\REFA}[1]{{\textcolor{black}{#1}}}
\newcommand{\REFB}[1]{{\textcolor{black}{#1}}}

\shorttitle{Reduced Representations of Turbulent Rayleigh-Bénard Flows via Autoencoders}
\shortauthor{M. Vinograd and P. Clark di Leoni}

\title{Reduced Representations of Rayleigh-Bénard Flows via Autoencoders}
\author{Melisa Y. Vinograd\aff{1}\aff{,2}
  \corresp{\email{mvinograd@udesa.edu.ar}},
 \and P. Clark Di Leoni\aff{1}\aff{,3}}

\affiliation{\aff{1}Universidad de San Andrés, Buenos Aires, Argentina
\aff{2}Departamento de Física, Universidad de Buenos Aires, CABA, Argentina
\aff{3}CONICET, Argentina
}

\begin{document}
\maketitle

\begin{abstract}
We analyzed the performance of Convolutional Autoencoders in generating reduced-order representations \REFA{the temperature field} of 2D Rayleigh-Bénard flows at $\Pr=1$ and Rayleigh numbers extending from $10^6$ to $10^8$, capturing the range where the flow transitions to turbulence.
We present a way of estimating the minimum number of dimensions needed by the Autoencoders to capture all the relevant physical scales of the \REFA{data} that is more apt for highly multiscale flows than previous criteria applied to lower dimensional systems.
We compare our architecture with two regularized variants as well as with linear methods, and find that manually fixing the dimension of the latent space produces the best results.
We show how the estimated minimum dimension presents a sharp increase around $\Ra\sim 10^7$, when the flow starts to transition to turbulence. Furthermore, we show how this dimension does not follow the same scaling as the physically relevant scales, such as the dissipation lengthscale and the thermal boundary layer.
\end{abstract}


\section{Introduction}\label{sec:introduction}

Generating reduced-order representations of turbulent flows is a major challenge underpinning several tasks and goals in fluid dynamics. They are essential both for modeling, as when writing Reduced-Order Models, and for analysis and theory, as, for example, in the quest to determine the existence, dimensions and characteristics of inertial manifolds.
Thus, the list of techniques for dimensionality reduction and their associated history is long, going from Fourier decompositions and Principal Component Analysis, to wavelets, and lately to neural network-based Autoencoders~\citep{bruntonBook}. While the former employ linear techniques to derive orthogonal bases to best represent the data, the latter perform non-linear decompositions and have been shown to be significantly more effective at generating compact representations of different datasets \citep{hinton_reducing_2006,GoodBengCour16}.

Fueled by their achievements in various machine learning tasks, autoencoders have been applied to various problems in fluid mechanics. In the context of the Kuramoto-Shivashinky equation, \cite{linot_deep_2020, linot_data-driven_2022} employed a combination of principal component analysis and autoencoders to parameterize the inertial manifold of the equation, where the dynamics on the parametrized space were later modeled with Neural ODEs \citep{chen_neural_2019}. These processes served to estimate the dimension of the inertial manifold, matching the estimates obtained via analysis of unstable periodic orbits (UPOs) \citep{ding_estimating_2016}.
Expanding on these methods,~\cite{floryan_data-driven_2022} incorporated elements of manifold theory into their analysis of the latent space.
Working with the Kolmogorov flow,~\cite{page_revealing_2021,page_exact_2023} used autoencoders, along with a Fourier decomposition of the latent space, to find and characterize UPOs and relate them to bursting events. Similarly, \cite{linot_dynamics_2023}, \cite{dejesus2023} and \cite{constante2024} used models of the latent space dynamics to search for UPOs in the flow in Kolmogorov, Channel and Pipe flows, respectively.
\cite{fukami_convolutional_2020} and \cite{,eivazi_towards_2022} were able to determine a hierarchy of modes within latent space and facilitated the proposal of nonlinear mode decomposition of flows in fewer modes than their linear counterparts.
In the realm of turbulent stratified flows, \cite{foldes_low-dimensional_2024}  were able to improve the autoencoders capabilites to capture high-order statistics by adding extra penalizations to the loss functions. 

In order to avoid the costly parameter sweeps involved in determining the smallest dimension needed for the autoencoder to represent the data properly, different regularized autoencoders variants have been proposed.
\cite{zeng_autoencoders_2023} utilized the Implicit Regularization method, first introduced by \cite{jing_implicit_2020}, to Kolmogorov flow, showing that, in their examples, the regularized variant was able to give the same estimate for the minimal numbers of modes as the non-regularized version. \cite{sondak_learning_2021} introduced $L_1$ regularization to the autoencoder model so as to create a sparse version of the latent vector. They successfully identified the optimal reduced latent space for the Kuramoto-Shivashinkyy equation while demonstrating the absence of a corresponding reduced latent space for the Korteweg-de Vries equation, reflecting its inherently infinite-dimensional dynamics. These regularized variants have not yet been applied to flows with large scale separation.

As was mentioned previously and as many of the examples above show, autoencoders serve as a powerful tool to generate parametrizations of inertial manifolds (or, at the very least, parametrizations of submanifolds that capture the relevant dynamics) and estimate their dimensions. Inertial manifolds are finite-dimensional manifolds that contain the global attractor of nonlinear dissipative systems~\citep{hopf_mathematical_1948,foias_inertial_1988}, offering a simplified description through a finite set of ordinary differential equations~\citep{temam_inertial_1990}. This concept has been successfully applied to systems such as the Kuramoto-Sivashinsky equation, where the existence of such manifolds has been rigorously established~\citep{hyman_kuramoto-sivashinsky_1986}. While this is not the case for Navier-Stokes flows~\citep{temam_inertial_1990}, several studies provide evidence of its possible existence~\citep{chandler2013a,willis_symmetry_2016,crowley2022,dejesus2023}. 

The main aim of this paper is to generate low-order representations of \REFA{the temperature field of} Rayleigh-Bénard flows and use them to estimate the \REFA{dimensionality of the flow of the flow as it transitions to turbulence}, with an added emphasis on understanding the performance of different autoencoders when dealing with strongly turbulent and multiscale data. Heat-induced currents play a crucial role in various geophysical and industrial activities. For instance, in the geophysical context, these currents are essential for the circulation of the Earth's atmosphere~\citep{hartmann_tropical_2001} and oceans~\citep{vage_ocean_2018}. The presence of buoyancy enables the generation of convective rolls, large-scale structures that transport heat and matter from hot to cold regions of the fluid. The balance between buoyant and viscous forces weighted by the Prandlt number, which is the ratio of viscosity to thermal diffusivity, is measured by the Rayleigh number. As this balance changes, so does the behavior of the flow. We focus on Rayleigh numbers ranging from $10^6$ to $10^8$, where turbulence starts to develop and the system experiences a sudden change in its properties, such as the emergence of multi-stability and long transient states~\citep{van_der_poel_flow_2012}, the reversal of large-scale structures \citep{sugiyama2010,castillo2019}, the appearance of new optimal steady states \citep{kooloth_coherent_2021}, and a loss of synchronization capabilities~\citep{agasthya_reconstructing_2022}. These changes are associated with an increase in the number of structures the flow exhibits, which, in turn, indicate that the dimension of the inertial manifold of the system is increasing during the transition. \REFA{The use of convolutional autoencoders to generate reduced-order models of Rayleigh-Bénard flows has been explored for $\Ra=10^7$ in~\cite{pandey_direct_2022}, where the local convective heat flux, a central measure for characterizing mean turbulent heat transfer, was forecasted by reducing the datasets to a dimension of 40.}

This article is organized as follows: Section~\ref{sec:methodology} details the theoretical background and the analytical methods employed, first introducing important aspects of Rayleigh-Bénard flows in Section~\ref{subsec:RB_flow}. This is followed by a detailed explanation of the different autoencoder architectures used in our study, \ref{subsec:FdAE} outlines the design and functionality of the \mbox{Fixed-$d$~Autoencoder} (F$d$AE) networks with a fixed latent space dimension, and~\ref{subsec:variable_d_AE} introduces two modifications to the network to improve training times: IRMAE and SIAE. In~\ref{subsec:numerical_simulations} we outline the empirical framework used to test our theories, the details of the numerical simulations and training of the neural networks. Results and conclusions are presented in Sections~\ref{sec:analysis} and~\ref{sec:conclusions}, respectively. 

\section{Problem set-up and methodology}\label{sec:methodology}

\subsection{Rayleigh-Bénard flow}\label{subsec:RB_flow}

Rayleigh-Bénard convection describes the motion of a planar cell of fluid flow that is heated from the below. Under the Boussinesq approximation, density variations are assumed to be small and the velocity and temperature fields are coupled through the buoyancy term. For a rectangular domain of horizontal length $L$ and height $h$ the equations take the form

\begin{align}
\frac{\partial \bm{v}}{\partial t} + (\bm{v} \cdot \nabla) \bm{v} &= \bm{\nabla} p + \nu \nabla^2 \bm{v} - \alpha g T \mathbf{\hat{z}}, \label{eqs:RB_vel} \\
\bm{\nabla} \cdot \boldsymbol{v} &= 0,  \label{eqs:RB_incomp} \\
\frac{\partial T}{\partial t}+(\bm{v} \cdot \boldsymbol{\nabla}) T &= \kappa \nabla^2 T, \label{eqs:RB_temp}
\end{align}
where $\bm{v}$ is the velocity field, $p$ the pressure, $T$ the temperature, $\nu$ the kinematic viscosity, $\kappa$ the thermal conductivity, $\hat{\boldsymbol{z}}$ the direction parallel to gravity and $g$ the acceleration of gravity. For reasons outlined below, we work with the rescaled temperature fluctuations $\theta = (T - T_0) \sqrt{\alpha g h / \Delta T}$, with $T_0 = T_b - z \Delta T/h$ the linear background profile and $T_b$ the temperature of the bottom plate. We assume periodic boundary conditions in the horizontal direction and non-slip boundary conditions in the vertical direction throughout the whole work. 

The dimensionless Rayleigh number, $\Ra$, quantifies the balance between buoyancy forces and the combined effects of thermal diffusion and viscous dissipation and is given by
\begin{equation}
\mathrm{Ra} = \frac{\alpha g h^3 \Delta T}{\nu \kappa}.
\label{eq:Ra}
\end{equation}
As $\Ra$ increases, the fluid undergoes several transitions: from stable stratification, characterized by conductive heat transfer, to the emergence of convective rolls or cells (Bénard cells), and eventually to chaotic and turbulent motion. At $\Ra \approx 1700$ the fluid transitions from conductive to convective behavior. This occurs as linear instability mechanisms break the translational invariance of the flow leading to the appearance of convective rolls. We focus on $\Ra$ values ranging from $10^6$ to $10^8$. At these levels, the dynamics are highly convective, with rolls starting to break down and the flow showing clear signs of turbulence. A subsequent regime, not covered in this work, occurring when $\Ra$ reaches $10^{13}$, is the so-called ultimate regime. Here, small-scale structures dominate, and rolls are no longer evident \citep{zhu_transition_2018}.

The two other dimensionless numbers that characterize Rayleigh-Bénard convection are the Prandtl number $Pr = \nu / \kappa$, which we set to $1$ in this study, and the Nusselt number $\Nu$

\begin{equation}
\Nu(z)=\frac{\left\langle  v_z T - \kappa \partial_z T   \right\rangle_{x, t}}{\Delta T \, \kappa / h},
\label{eq:nusselt}
\end{equation}
where $\langle \cdot \rangle{x,t}$ is the time and spatial average at a height $z$. The Nusselt number compares the ratio of vertical heat transfer due to convection to the vertical heat transfer due to conduction. On account of energy conservation, the Nusselt number is constant at each height $z$ when using periodic or adiabatic boundary conditions on the sides of the domain.

In terms of lengthscales, there are two important small-scale quantities. One is the thermal boundary layer $\delta_\theta$, which is the thin region adjacent to the walls where most of the temperature gradient of the fluid is concentrated. In this layer, temperature changes significantly, while beyond it, the fluid temperature becomes comparatively uniform.  Understanding how $\delta_\theta$ scales with varying Rayleigh and Nusselt numbers gives insight into the dynamics of convective heat transfer. The thermal boundary layer thickness can be estimated by identifying the height at which the root-mean-square (rms) of the temperature deviates from linear behavior near the plates.

The other lengthscale is the Kolmogorov scale $\eta_K(\bm{x},t)$, which represents the size at which viscous forces damp out the smallest turbulent eddies and is given by~\citep{Scheel_2013}

\begin{equation}
    \eta_K(\bm{x}, t)= \left( \frac{\nu^3}{\varepsilon(\bm{x}, t)} \right)^{1/4},
\end{equation}
where $\epsilon(\bm{x}, t)$ is the kinetic energy dissipation rate which is defined as
\begin{equation}
\epsilon(\bm{x}, t) = \nu \frac{1}{2} (\nabla \bm{v} + \nabla \bm{v}^T)^2.
\end{equation}
Due to the symmetries of the problem, $\eta_K$ should be homogeneous in time and the horizontal direction but not in the vertical direction. Further details on how this behave are given below when the datasets used are described.

Finally, the characteristic timescale of the system is the large eddy turnover time, denoted by $\tau$. This is calculated as the time required for a large-scale eddy to complete one full rotation over the height of the domain $\tau = 2h / w$, with  $w$ the root mean square of the vertical velocity component, \(w \equiv v_{z, \text{rms}}\).

\subsection{Fixed-$d$ Autoencoders}\label{subsec:FdAE}

Autoencoders are a type of dimensionality reduction algorithms that seek to construct an encoding function $\mathcal{E}$ that can map an input field $u \in \mathbb{R}^{N_x \times N_z}$ into a smaller-sized latent vector $\bm{z}$ and a decoding function $\mathcal{D}$ that can perform the inverse operation. This is done by approximating $\mathcal{E}$ and $\mathcal{D}$ with neural networks and training the resulting model $\mathcal{F}(u) = \mathcal{D}(\mathcal{E}(u))$ to approximate the identity operation $u = \mathcal{F}(u)$, where the latent vector $\bm{z}=\mathcal{E}(u)$ acts as a bottleneck layer. The training process involves minimizing a prescribed loss function
$\mathcal{L}(u, \mathcal{F}(u))$ with respect to the parameters of the neural networks. The most common choice of loss function (and the one used in the majority of this work) is the $L_2$ norm

\begin{equation}
\mathcal{L}(u, \tilde{u}) =
\frac{1}{B} \sum_{b=1}^{B} L_2(u^{(b)} - \tilde{u}^{(b)}) = \frac{1}{B} \sum_{b=1}^{B}
\sum_{i=1}^{N_x} \sum_{j=1}^{N_z} \left( u_{ij}^{(b)} - \tilde{u}_{ij}^{(b)} \right)^2,
\label{eq:L2}
\end{equation}
where $\tilde{u}^{(b)} = \mathcal{F}(u^{(b)})$, and $u_{ij}^{(b)}$ and $\tilde{u}_{ij}^{(b)}$ are the individual grid elements of each field and $B$ denotes the batch size, representing the number of training samples processed simultaneously during each iteration of the training.

\begin{figure}
  \centerline{\includegraphics[width=0.8\linewidth]{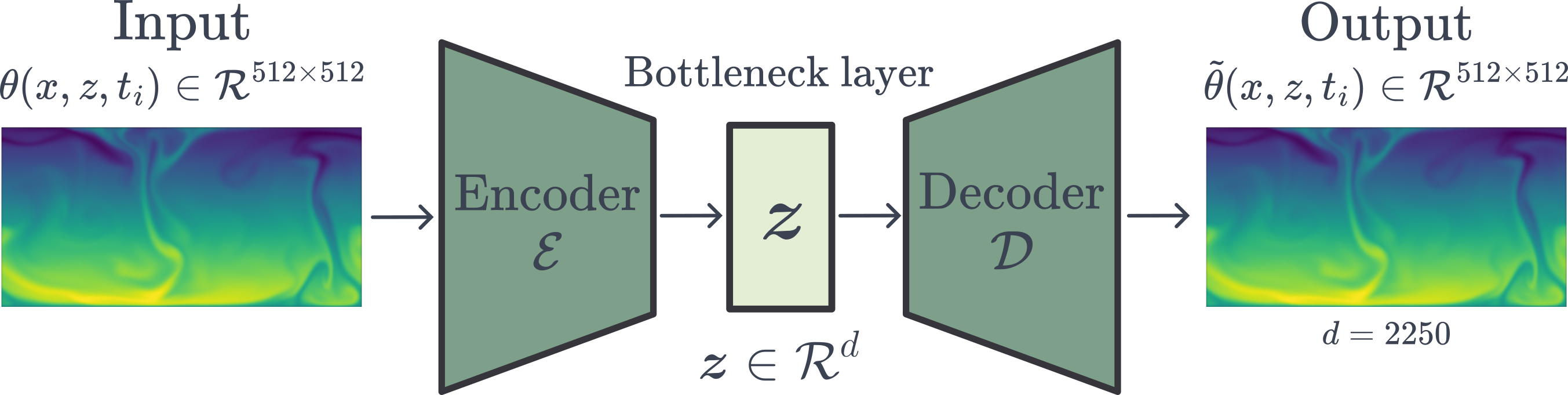}}
  \caption{Two-dimensional convolutional neural network-based autoencoder. The central layer $z$ in the middle is the bottleneck layer or latent vector. The image on the left side represents an example of the input to the network for $\Ra=10^8$, corresponding to the temperature fields of a Rayleigh-Bénard numerical simulation. The image on the right is the output of the network for the same input, with a latent space dimension of $d=2250$.}
\label{fig:cnn_architecture}
\end{figure}

\begin{figure}
  \centerline{\includegraphics[width=0.8\linewidth]{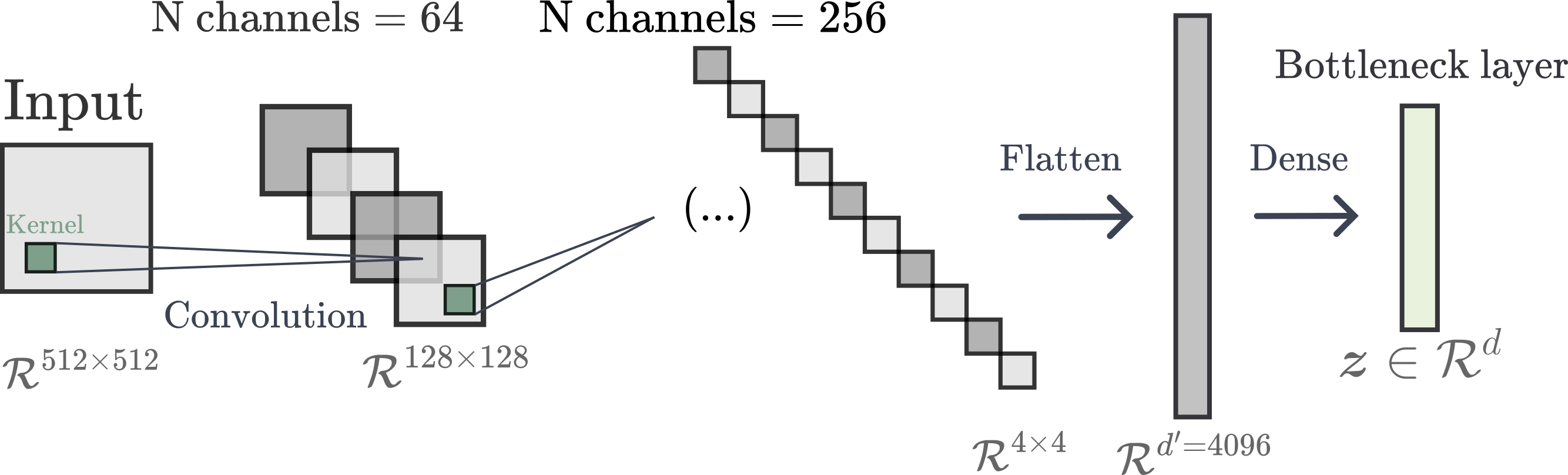}}
  \caption{Schematic of the convolutional encoder $\mathcal{E}$. The first layers are convolutional while the last layer after the flattening operation is a fully-connected layer. Details are listed in table~\ref{tab:cnn_parameters}.}
  \label{fig:cnn_specific}
\end{figure}

A diagram of the autoencoders used is shown in figure~\ref{fig:cnn_architecture}. Our input data are snapshots of the rescaled temperature field $\theta(x, z, t_i)$ and the output is denoted as $\tilde{\theta}$.  For simplicity, we also define $\Delta \theta = \theta - \tilde{\theta}$. We used convolutional neural networks for both the encoder and decoder as is standard in image-processing tasks. A schematic of the encoder architecture is shown in figure~\ref{fig:cnn_specific}. The input fields of size $\mathcal{R}^{512 \times 512}$ undergo transformations through four convolutional layers, each employing a $5 \times 5$ kernel (highlighted in green in the figure). The configuration of each subsequent layer is characterized by its shape, which includes the number of channels and the new dimension size. The reduction in size at each layer is achieved through the application of specific strides and padding. After the final convolutional layer, the output is flattened to maintain its structure which is a vector of dimension $d'$, followed by the introduction of a dense layer with a smaller dimension $d$. This additional layer serves to form the compact encoded representation of the input, known as the bottleneck layer. The different hyperparamters used are listed in table~\ref{tab:cnn_parameters}.
Note that under this set-up the dimension of the latent vector $d$ is a hyperparameter that has to be fixed prior to training. The primary goal of this work is to \REFA{evaluate} the performance of the autoencoder as we vary $d$ \REFA{in order to assess what is the minimum $d$ needed by the autoencoder to properly represent the data, in this case the temperature field}. We refer to this architecture as Fixed-$d$ Autoencoder, F$d$AE, variations are discussed below. Details about the training protocol and datasets used are given below. Note that for the value of $d'$ we estimated the minimum number of degrees of freedom necessary to represent the flow by determining the number of Fourier and Chebyshev modes needed to reconstruct the relevant small-scale dynamics. In the periodic x-direction, we need at least $2 k_\eta$ modes. For the non-periodic z-direction, we can calculate the number of interpolation Chebyshev points needed to resolve the boundary layer. This estimation indicates that at $\Ra \approx 10^8$ approximately 4096 modes are required (or degrees of freedom), thus determining the largest size (biggest bottleneck layer) to which we compressed our data using the networks.


\begin{table}
\begin{center}
\begin{tabular}{lc}
Input shape & $512 \times 512$  \\
Encoder structure & $64: 64: 128: 256$ \\
Decoder structure & $256: 128: 64: 64$ \\
Kernel size & $5 \times 5$ \\
Strides & $4: 4: 4: 2$ \\
$d'$ & $4096$ \\
Activation function & ReLU \\
Last activation function & Tanh \\
$L_1$ regularization & $\lambda \in [10^{-3}, 10^{-1}]$ \\
Implicit regularization & $n \in [1,4]$ \\
\end{tabular}
\caption{CNN Autoencoder Parameters. The encoder and decoder structure specify the number of filters per convolutional layer.}
\label{tab:cnn_parameters}
\end{center}
\end{table}

\subsection{Variable-$d$ Autoencoders}\label{subsec:variable_d_AE}

In the F$d$AE outlined above, the dimension of the latent space $d$ has to be chosen prior to training and a new neural network has to be trained for each latent space $d$. We now discuss two regularized variations where $d$ can be adjusted after training. In both variations, the networks are trained using an over-parametrized bottleneck layer of size $d'\gg d$ of which only $d$ parameters are used for inference. We then always use $d$ to refer to the effective dimension of the encoding.


\subsubsection{Sparsity-Inducing Autoencoders}\label{subsubsec:SIAE}

The first approach involves applying a $L_1$ regularization term, acting over $\bm{z}$, to the loss function \eqref{eq:L2}, resulting in 
\REFA{
\begin{equation}
 \mathcal{L}_S(u, \tilde{u}, \bm{z}) = \mathcal{L} + \frac{1}{B} \sum_{b=1}^{B} \lambda L_1(\bm{z}^{(b)}), 
\label{eq:total_loss_L1}
\end{equation}}

where $\mathcal{L}$ is defined in \ref{eq:L2} and $B$ denotes the batch size. The regularization term is defined as
\begin{equation}
L_1(\bm{z})=\frac{1}{d'} \sum_{j=1}^{d'} \left| z_j \right|,
\label{eq:L1}
\end{equation}
$\lambda$ is an extra hyperparameter which controls the intensity of the regularization, with $\lambda=0$ recovering the previous case.

By introducing this regularization, the model is encouraged to maintain only the most essential connections and promote sparsity in the middle layer. This approach was explored in~\cite{sondak_learning_2021}, where the authors employ an autoencoder with latent space penalization. We refer to this variation as SIAE, for Sparsity-Inducing Autoencoders.

As stated above, a SIAE is trained using a bottleneck layer of size $d'$. After training, the respective latent representation $\bm{z}'$ of each element of the training dataset is calculated, forming a matrix $Z'$, from which then the interquartile range (IQR) of each dimension is obtained. Finally, a sparse version of the latent vector $\bm{z}'$ is constructed where the $d'-d$ dimensions with the lowest IQR values are set to zero (and thus only $d$ non-zero dimensions remain). This whole operation can be viewed as replacing the encoder $\mathcal{E}$ by
\begin{equation}
\mathcal{E}_S (\bm{x}) = \Theta_d (\mathcal{E}(\bm{x})),
\end{equation}
where $\Theta_d$ is the operator that sets the least import dimensions to zero, during inference, meaning, too, that the latent vector takes the form $\bm{z} = \Theta_d (\bm{z}')$. The decoder is unchanged in this version.
The structure of $\mathcal{E}$ and $\mathcal{D}$ themselves is the same as in the F$d$AE, except for the fact that the bottleneck layer has size $d'$, of course, this is true in the next version as well. The values of $d'$ and $\lambda$ are presented in table~\ref{tab:cnn_parameters}.

\subsubsection{Implicit Rank Minimizing Autoencoders}\label{subsubsec:IRMAE}

The second approach is known as implicit regularization (Implicit Rank Minimizing autoencoders, IRMAE)~\citep{zeng_autoencoders_2023}. This method uses the fact that adding depth to a matrix factorization enhances an implicit tendency towards low-rank solutions to construct low-rank versions of the latent representations~\citep{jing_implicit_2020}. It works by adding $n$ linear layers, $\mathcal{W}=\mathcal{W}_1, \mathcal{W}_2, \cdots, \mathcal{W}_n$, between the encoder and the bottleneck layer.

Similar to the SIAE, an IRMAE is trained with bottleneck layers of size $d'$. After training, the covariance matrix of the encodings $Z'$ is formed, but now instead of calculating the IQR values, a singular value decomposition, $Z'=U\Sigma U^*$, is performed. The final latent vector is obtained by projecting into the first $d$ dimensions. The resulting encoder used during inference has the form
\begin{equation}
\mathcal{E}_{I}(\bm{x}) = \Theta_d (U^* \mathcal{W} \mathcal{E}(\bm{x})),
\label{eq:IRMAE_encoder}
\end{equation}
where $\Theta_d$ truncates the singular values to those who are different from zero so that $U^* \bm{z}' \approx \Theta_d (U^* \bm{z}')$ holds.
The decoder, which now needs to project back into $Z'$, takes the form
\begin{equation}
\mathcal{D}_{I}(x) = \mathcal{D}(U \mathcal{E}_{I}(\bm{x})).
\label{eq:IRMAE_decoder}
\end{equation}
Contrary to~\cite{zeng_autoencoders_2023}, we keep the loss function the same as in the F$d$AE (equation~\eqref{eq:L2}) and do not add extra terms as we did not find that they changed the final results.

\subsection{Numerical simulations, datasets and training protocol}\label{subsec:numerical_simulations}

Equations~\eqref{eqs:RB_vel} to~\eqref{eqs:RB_temp} where solved using SPECTER~\citep{fontana_fourier_2020}, a highly-parallel pseudo-spectral solver which utilizes a Fourier-continuation method to account for the non-periodic boundary conditions in the vertical direction.
For implementation reasons, the code solves for the rescaled temperature $\theta$, instead of the temperature $T$. So, as explained above, we present our results in terms of $\theta$, this does not pose a problem for our analysis as the temperature fluctuations and the rescaled temperature fluctuations only differ by the scaling factor.
We solved the equations at 7 different Rayleigh numbers, ranging from $10^6$ to $10^8$. In all cases, the Prandtl number was set to $Pr = 1$, and the domain sizes were $[L_x, L_z] = [2 \pi, \pi]$, with $N_x = N_z = 512$ grid points for spatial resolution in each direction. This spatial domain ensured that the smallest scales were resolved for all values of $\Ra$. The values of the viscosity used at each $\Ra=\{10^6,\; 5\times10^6,\;  7.5\times10^6; 10^7,\; 1.75\times10^7,\; 2.5\times10^7,\; 5\times10^7,\; 10^8 \}$ were $\nu=\{9.86 \times 10^{-3},\; 4.41 \times 10^{-3},\; 3.60 \times 10^{-3},\; 3.12 \times 10^{-3},\; 2.36 \times 10^{-3},\; 1.97 \times 10^{-3},\; 1.39 \times 10^{-3},\; 9.68 \times 10^{-4} \}$,  respectively.
Simulations were started from random initial conditions for the temperature field, allowed to reach a statistically stationary state and then run for over 700 turnover times. For each $\Ra$ value, we conducted two simulations with different initial conditions.

\begin{figure}
    \centering
    \includegraphics[width=\linewidth]{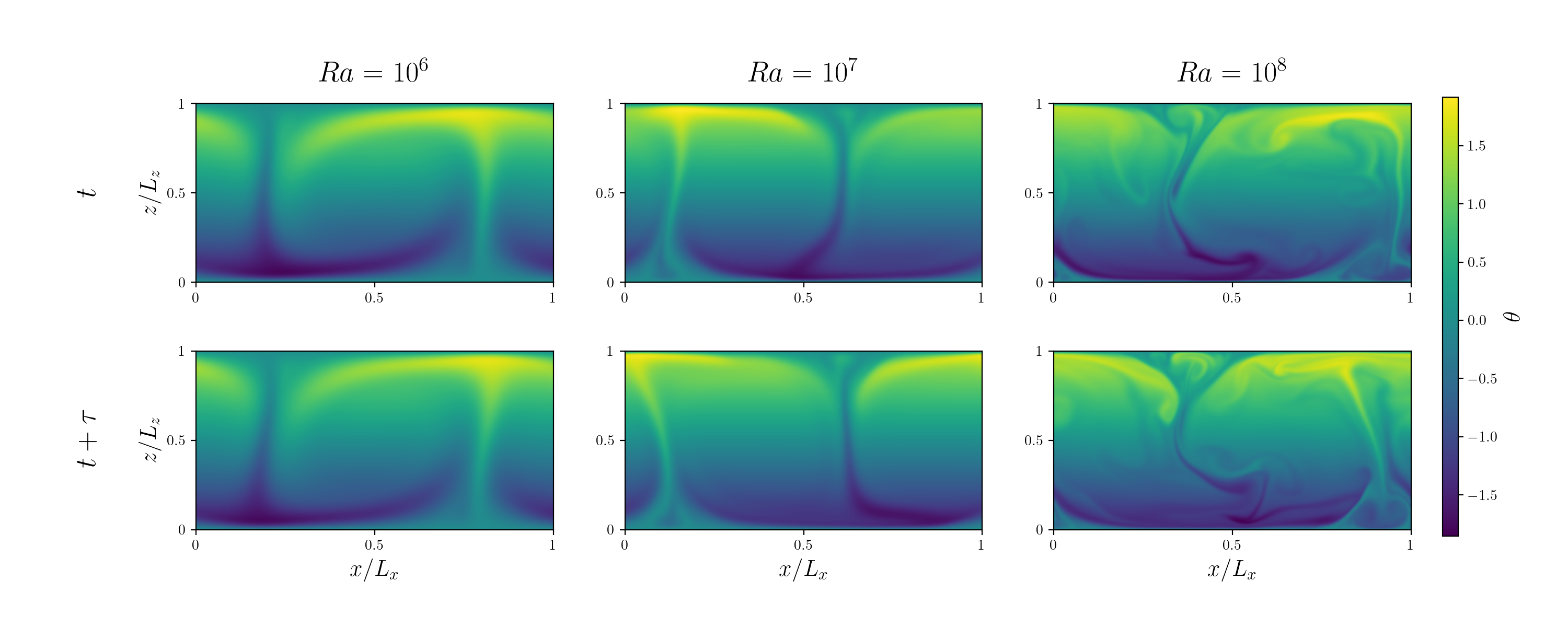}
    \caption{Visualizations of $\theta(x,z,t)$ and $\theta(x,z,t+\tau)$ for three different values of $\Ra$.}
    \label{fig:snapshots_of_ras}
\end{figure}

\begin{figure}
    \centering
    \includegraphics[width=0.8\linewidth]{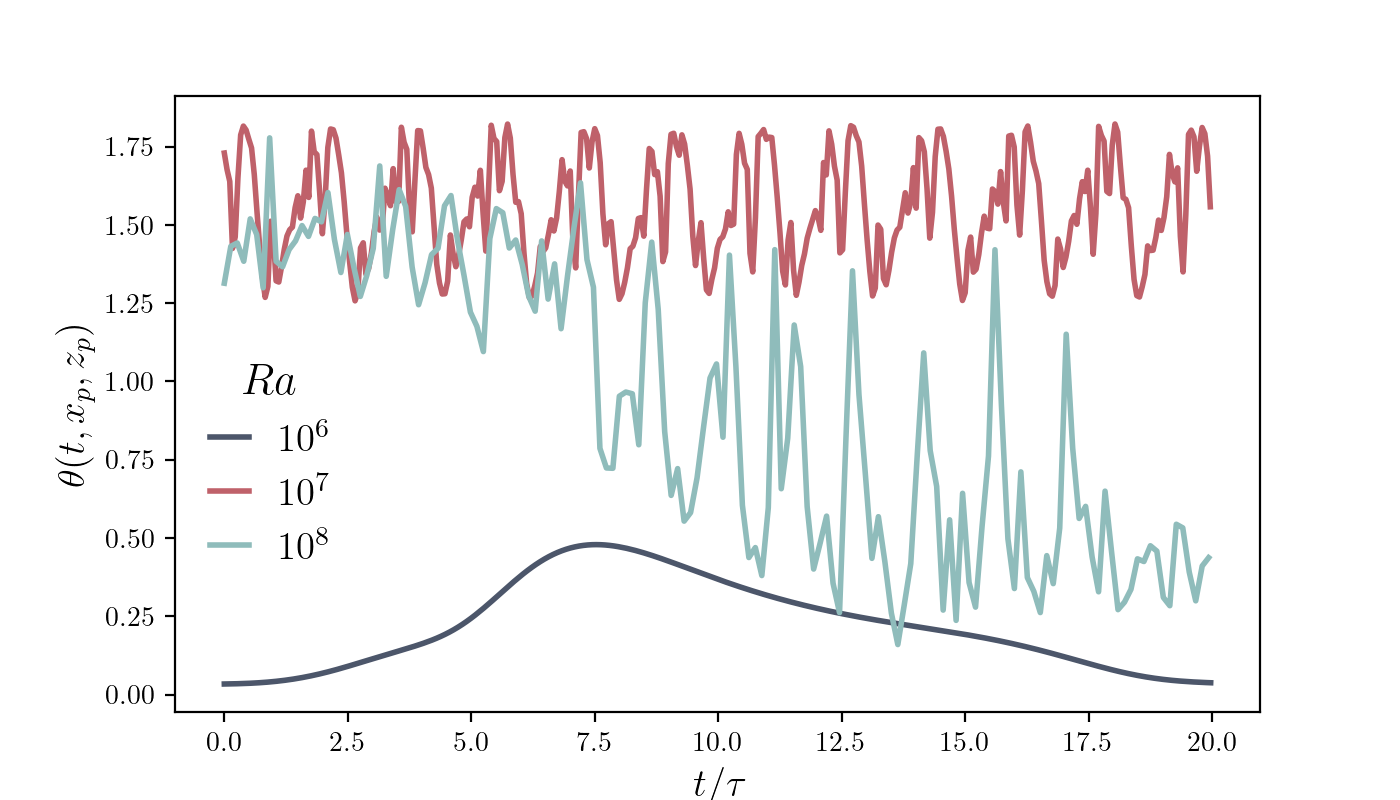}
    \caption{Temporal evolution at a fixed location $(x_p, z_p)$ of the flow for three different $\Ra$. }
    \label{fig:T_over_time}
\end{figure}

\begin{figure}
    \centering
    \includegraphics[width=0.8\linewidth]{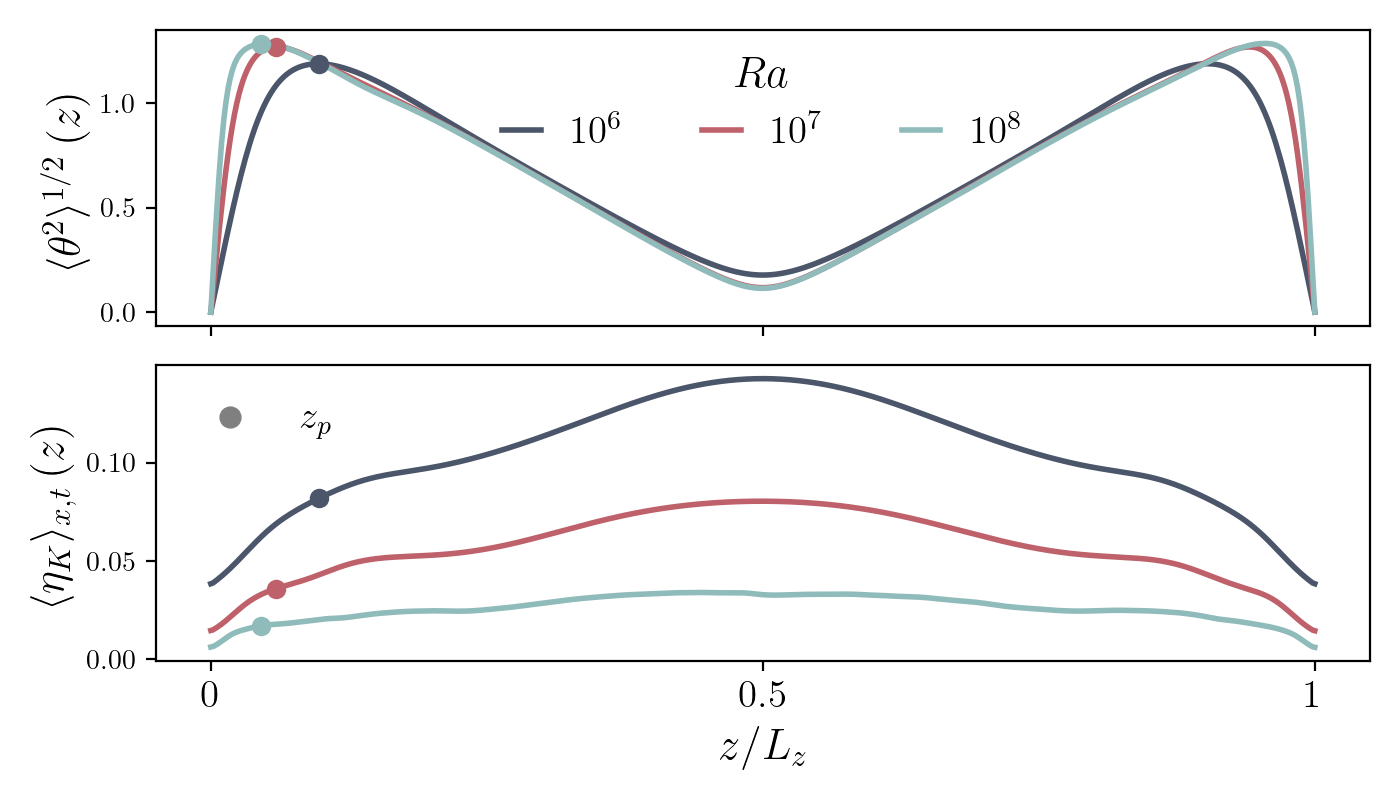}
    \caption{Relevant scales of the fluid for three different $\Ra$. The top plot illustrates the profile of $\langle \theta^2 \rangle^{1/2}_{x,t}$. The lower plot shows the Kolmogorov scale, $\langle \eta_K \rangle_{x,t} (z)$. For each $\Ra$, $z_p$ is marked.}
    \label{fig:eps_eta_theta}
\end{figure}

We now take a moment to describe three paradigmatic cases, $\Ra = 10^6$, $10^7$, and $10^8$. Figure~\ref{fig:snapshots_of_ras} shows snapshots of $\theta(x,y)$ at a time $t$ and one turnover time later, at a time $t+\tau$. As the Rayleigh number increases the flow becomes more complex with smaller and smaller structures popping up. New timescales appear in the flow too, as can be in seen in figure~\ref{fig:T_over_time}, where we show the temporal evolution of $\theta$ at one single point $(x_p, z_p)$ throughout several turnover times (while the location $x_p$ is indistinct due to the translational symmetry of the flow, the significance of $z_p$ will be discussed shortly). \REFA{The turnovertime $\tau$ is defined for each \Ra.} Even at $\Ra=10^7$ the flow exhibits multi-timescale behaviour and intricate dynamics. \REFA{The shift observed at $t=7.5\tau$ in the $\Ra=10^8$ case corresponds to a horizontal displacement of the plume.} Finally, we show profiles of the horizontally- and time-averaged RMS fluctuations, $\langle \theta^2 \rangle_{x,t}^{1/2}(z)$, and Kolmogorov lengths scale, $\langle \eta_K \rangle_{x,t} (z)$, in figure~\ref{fig:eps_eta_theta}. As the Rayleigh number increases, the thermal boundary layer becomes steeper, as expected. As the dissipation is not uniform in the vertical direction, the Kolmogorov lengthscale increases when approaching the middle of the domain. We are interested in checking how well the autoencoders represent each scale of the flow, for this matter we define $z_p(\Ra)$ \REFA{as the the height where the temperature perturbation fluctuations are highest. Below $z_p$ the dynamics are dominated by the boundary layer, while above the Kolmogorov lengthscale decreases when approaching the center of the domain.}   
The corresponding $z_p$ are marked in figure~\ref{fig:eps_eta_theta}. As we will explain below, corroborating if an autoencoder is properly representing this scale will be the benchmark used to test its capabilities and choose the minimal number of dimensions $d^*$ needed for the latent space dimension. As a summary, we list the values of $\eta_K (z_p)$, $\Rey$, $\Nu$, and $\delta_\theta$ for the three Rayleigh numbers cases just showcase in Table~\ref{tab:parameters}.

\begin{table}
  \begin{center}
    \def~{\hphantom{0}}
    \begin{tabular}{lccc}
      \Ra & $10^6$ & $10^7$ & $10^8$ \\[3pt]
      \hline
      $\eta_K (z_p)$ & 0.06 & 0.03 & 0.01 \\
      \Rey & 300 & 840 & 3300 \\
      $\Nu$ & 7.7 & 14 & 25 \\
      $\delta_\theta$ & 0.07 & 0.04 & 0.02\\
      \hline
    \end{tabular}
    \caption{Parameters used for the reference Rayleigh-Bénard flows, all in non-dimensional simulation units. $\Rey = u_{rms} h / \nu$ denotes the Reynolds number, where $u_{rms}$ is the root mean square velocity. }
    \label{tab:parameters}
  \end{center}
\end{table}

As input for the networks, we utilized the 2D $\theta$ fields from the numerical simulations described above, selecting specific snapshots for each dataset. We excluded the transient regime of the flow, as we are interested in the statistically steady states. \REFA{The cutoff point for the transient regime was defined as the point when the volume-averaged kinetic energy began to stabilize, oscillating around a constant value.} To address the continuous spatial symmetry in the horizontal direction, we augmented the data by applying 10 random shifts over the $x$-axis at each timestep. This augmentation increased the number of temperature fields under consideration from $7000$ to $70 \, 000$, spanning over 700 turnover times. Note that it would also have been possible to deal with this symmetry by performing phase-shifts, as described in~\cite{linot_deep_2020} and formulated in~\cite{budanur_reduction_2015}, instead of data augmentation, we chose to explicitly retain the symmetry as we are interested in exploiting it in future analysis \citep{page_revealing_2021,page_exact_2023}. The datasets were normalized with respect to the global maximum and minimum, in order to feed $\theta / \theta_0 \in[-1,1]$ to the neural networks. Additionally, a separate, longer dataset was created for testing to enable the integration of phenomena over extended periods, facilitating a more comprehensive evaluation of the performance of the model over longer time scales for certain physical metrics, such as the estimation of the Nusselt number.

During the training process, we employed the Adam optimizer with its default configurations to minimize the $L_2$ norm (equation~\ref{eq:L2}), which was also monitored on the validation set. Each network was trained over 500 epochs, by which point all networks had already stabilized. We monitored the validation metrics, and the batch size was set to 32. To ensure consistency, three different models were trained for each variation of the latent space dimension $d$. Each model was trained using different snapshots from the same extensive simulation, with both the training and test datasets maintained at the same size. \REFA{Specifically, for each model, we performed three independent runs: (1) using a random 80-20 train-test split, (2) reinitializing the network with different learning rates and initialization seeds but keeping the same data split, and (3) using the same initialization but applying a different random 80-20 train-test split. In this last case, we also introduced new random shifts to the data.} The results shown in the next section reflect the averaged values from these three variations.

\section{Results}
\label{sec:analysis} 


All results shown here are calculated over elements of the test datasets and averaged over them when appropriate.

\subsection{Fixed-$d$ Autoencoders}

\begin{figure}
  \centerline{\includegraphics[width=1\linewidth]{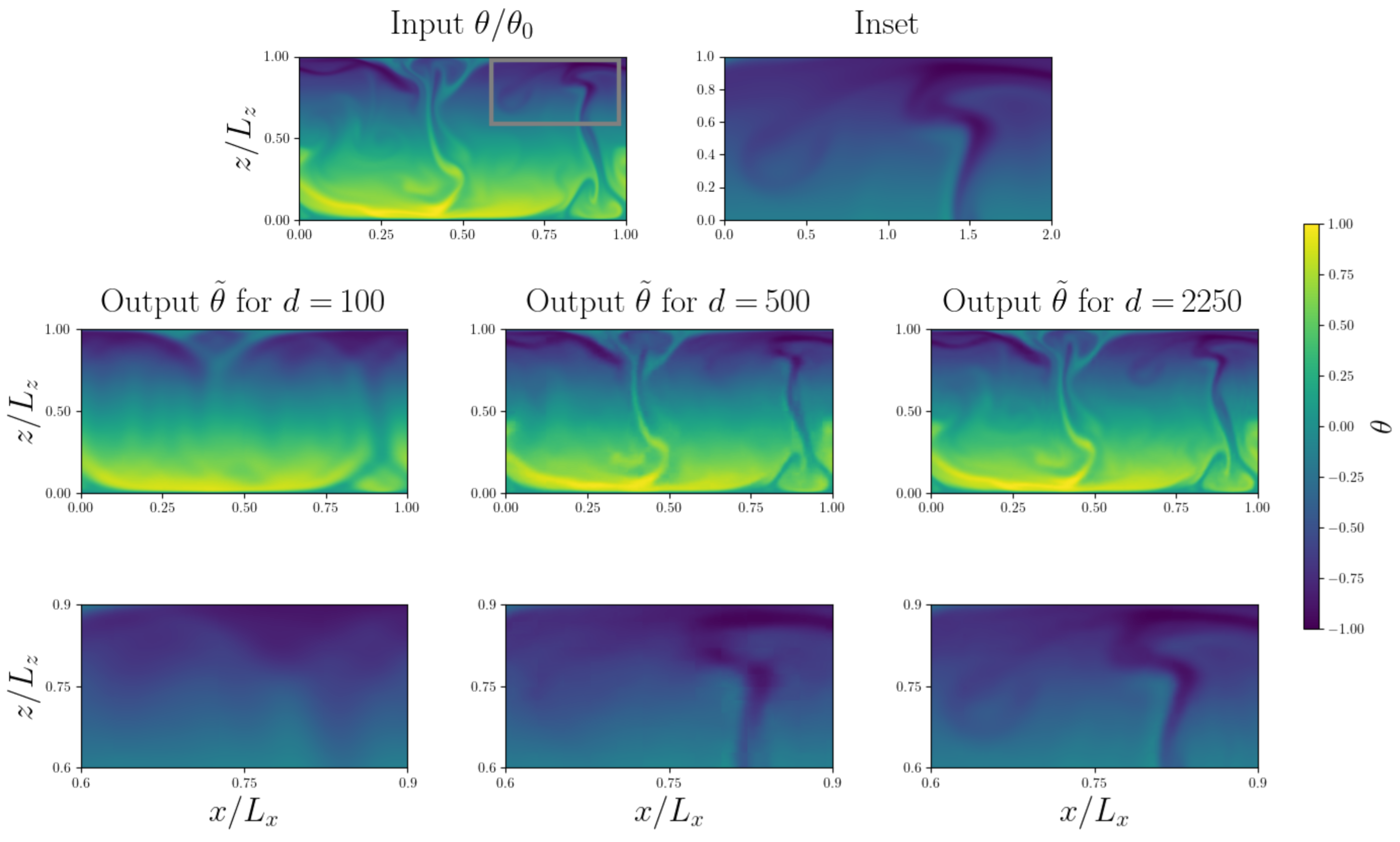}}
  \caption{The output fields $\tilde{\theta}$ of the F$d$AE for a snapshot of the temperature field, corresponding to latent space dimensions $d = 100$, 500, and 2250, with respective MSEs of $1.6 \times 10^{-3}$, $3.1 \times 10^{-4}$, and $4.0 \times 10^{-5}$, are shown. For comparison, the normalized ground truth field $\theta / \theta_0$, used as input, is also displayed. Additionally, an inset highlights a region where small-scale structures are present.}
\label{fig:visualizations}
\end{figure}

We focus our analysis on the performance of the F$d$AE on the $\Ra = 10^8$ case, as this is the highest Rayleigh number used in our study, and thus the most complex case. We reference results at other Rayleigh numbers when appropriate. At this particular value of $\Ra$ the flow is characterized by a rising hot plume and falling cold plume, which generate two large counter-rotating vortices which are accompanied by a rich variety of small-scale structures, the F$d$AE has to be able to capture the whole range of scales. In figure~\ref{fig:visualizations} we present a snapshot of the original temperature field $\theta$ (chosen out of the test dataset) together with its respective output $\tilde{\theta}$ obtained for selected values of latent space dimension $d$, along with zoomed-in views of regions with small-scale structures. These values—100, 500, and 2250—are noteworthy as they represent critical thresholds in performance and mark the transition points in the ability of the model to go from just reconstructing the large convective rolls to reproducing the small-scale details.

\begin{figure}
    \centering
    \includegraphics[width=0.8\linewidth]{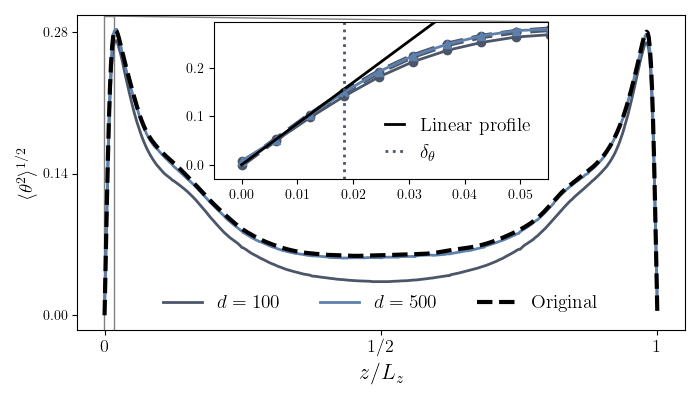}
    \caption{Profile of the original $\theta_{\text{rms}}$ along with reconstructed profiles for F$d$AE with $d=100$ and $d=500$. The inset zooms in on the near-wall region, showing the boundary layer thickness $\delta_\theta$, defined as the height $z$ where the $\theta_{\text{rms}}$ profile transitions away from a linear profile. While the reconstruction with $d=100$ does not capture $\theta_{\text{rms}}$ accurately in the bulk, the reconstruction with $d=500$ successfully reproduces the boundary layer behavior.}
    \label{fig:theta_rms}
\end{figure}

\begin{figure}
    \centering
    \includegraphics[width=0.8\linewidth]{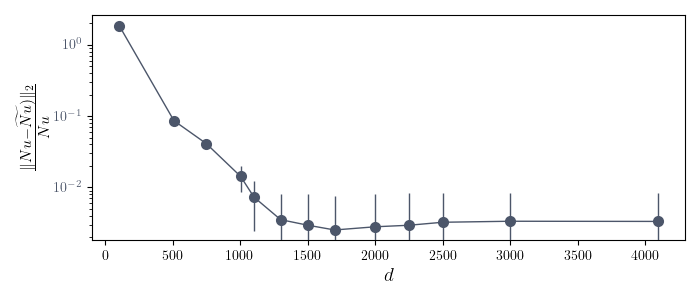}
     \caption{Difference between the Nusselt number $\Nu$ and the reconstructed Nusselt number $\tilde{\Nu}$ with F$d$AE as a function of the latent dimension $d$. The error stabilizes at $d=1500$, indicating that the model reliably captures $\Nu$ at this dimension. For reference, the true value of $\Nu$ is provided in Table 1.}
    \label{fig:nusselt_error}
\end{figure}

When $d=100$, the F$d$AE mainly reconstructs the broader flow patterns and larger convective rolls. Increasing the latent space dimension to $d=500$ enhances the performance of the autoencoder, which can now capture mid-range structures. \REFA{At this value of $d$, the model begins to accurately reproduce physically relevant quantities, such as the boundary layer thickness $\delta_\theta$. In figure~\ref{fig:theta_rms}, we show the profile of $\theta_{\text{rms}}$ along with its reconstructions for $d=100$ and $d=500$. The inset highlights the near-wall region where the boundary layer thickness $\delta_\theta$ is defined as the height $z$ at which the $\theta_{\text{rms}}$ profile departs from a linear profile. While the $d=100$ reconstruction does not accurately capture $\theta_{\text{rms}}$ in the bulk, the $d=500$ model successfully reproduces the boundary layer structure.} \REFA{
Finally, at $d=2250$, the autoencoder is visually able to express the finer details of the flow. To further quantify this, we examine the accuracy of $\Nu$ as a function of latent dimension $d$. In figure~\ref{fig:nusselt_error} shows the difference between the true Nusselt number $\Nu$ and the reconstructed Nusselt number $\widetilde{\Nu}$ across varying $d$. This difference stabilizes around $d=1500$, indicating that the model reliably captures $\Nu$ at this dimension.} \REFA{
These figures collectively serve as a visual testament to the power of the F$d$AE in identifying and reconstructing features across various scales, highlighting the importance of latent space dimensionality in the performance of the model. One of the main goals of this paper is to introduce a clear metric to determine the specific value of $d$ at which all relevant scales are accurately reconstructed, as discussed in detail below.
}

\begin{figure}
  \centerline{\includegraphics[width=0.85\linewidth]{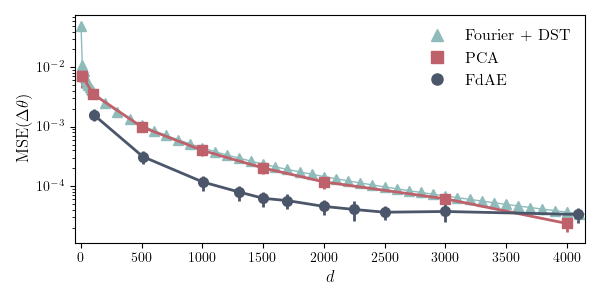}}
  \caption{Comparison of the MSE of the temperature field ($\theta$) reconstructions against the latent space dimension $d$ for the F$d$AE (represented by circles), PCA (illustrated with squares), and Fourier-DST decomposition (denoted by triangles) using the same dataset. For the autoencoder, each data point corresponds to the average of three runs with minor variations.}
\label{fig:L2}
\end{figure}

Figure~\ref{fig:L2} shows the mean squared errors between input and output, MSE$(\Delta \theta)$, as a function of latent space dimension, $d$. Additionally, the plot includes comparative lines representing the results of Principal Component Analysis (PCA) and a decomposition combining Fourier and Discrete Sine Transform (DST) in the $z$ (non-periodic) direction for different latent space dimensions, i.e., number of modes used in each case. Such comparisons are instrumental in benchmarking the performance of the F$d$AE against these established linear techniques. To ensure a fair comparison, when doing PCA we projected the training dataset and then evaluated the test data using the modes obtained from the initial decomposition. It is important to note that due to the large size of the training dataset matrix, which stacks all the flattened images, performing PCA directly is too computationally intensive. Consequently, we resorted to using Incremental PCA from Scikit-Learn~\citep{scikit-learn}, which is an suitable approximation method for large datasets that computes PCA over small batches. 

In neither case does the MSE exhibit a sharp drop-off that would indicate the minimum number of dimensions needed to accurate represent the flow. This is to be expected in systems with this degree of scale separation, for example, when studying the Kuramoto-Sivashinsky equation, \cite{linot_data-driven_2022} found sharp decreases in the MSE at the $d$ coinciding with that estimated for the inertial manifold only when using small domain sizes, when they increased the domain the drop-off became noticeably smoother. Similar results were observed in studies of turbulent pipe flows \citep{constante2024}. This effect is only greater in our case.
The F$d$AE outperforms the linear methods in terms of the MSE within the test dataset, demonstrating its capacity to capture the non-linearity attributes present in the dataset. Furthermore, it is important to acknowledge that the MSE might not be the most fitting metric to gauge the fidelity of reconstructions, particularly for smaller scales, as it predominantly captures the high energy modes, which are effectively resolved even at lower latent space dimensions as demonstrated in figure~\ref{fig:visualizations}. 

\begin{figure}
    \centerline{\includegraphics[width=0.9\linewidth]{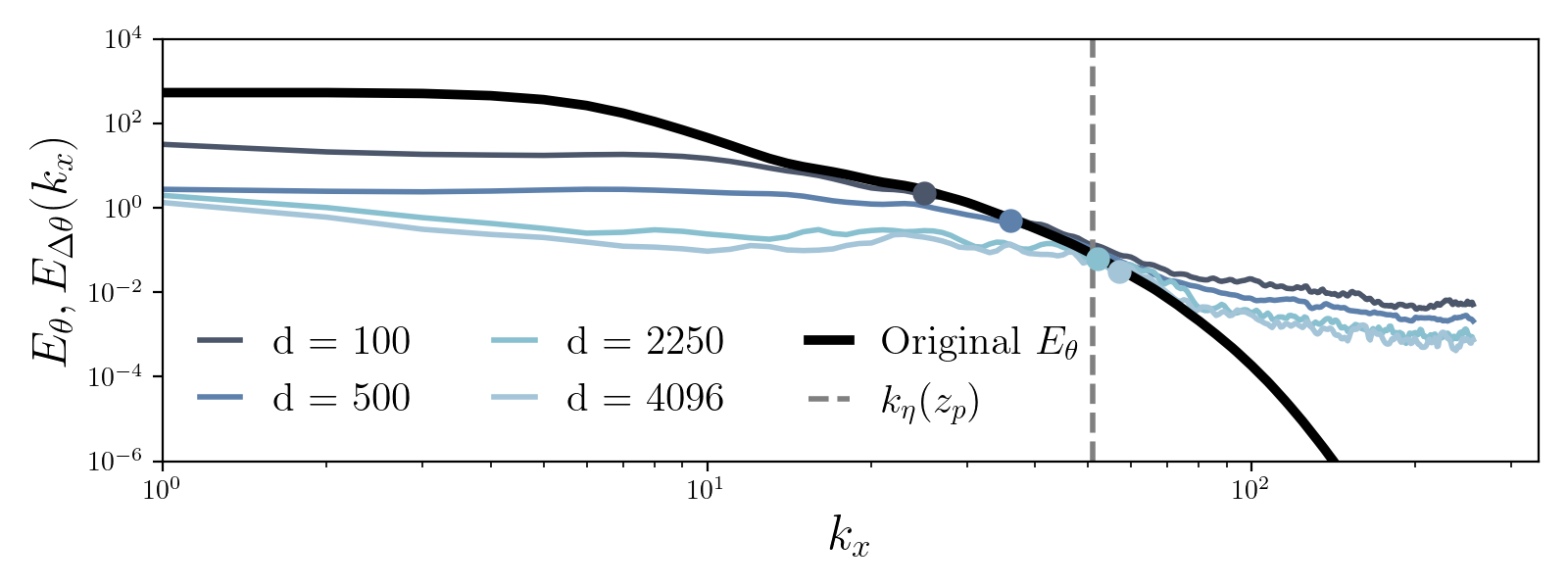}}
    \caption{Error spectra $E_{\Delta \theta}$ for various reconstructions, alongside the reference energy spectra $E_\theta$ for comparison. The spectra were calculated at $z_p$. The dotted vertical line indicates the wavenumber corresponding to the Kolmogorov scale $k_\eta$ also at $z_p$, and the dots represent the maximum resolved $k_x$ for each $d$.}
    \label{fig:error_spectra}
\end{figure}

To quantify the ability of the F$d$AE to represent different scales in the flow, we compute the energy spectrum of the difference between input and output
\begin{equation}
E_{\Delta \theta}\left(k_x\right)=\left\langle\left|\Delta \hat{\theta}\left(k_x, z_p, t\right)\right|^2\right\rangle_t,
\end{equation} 
where $\Delta \hat{\theta}\left(k_x, z_p, t\right)$ are the Fourier coefficients of $\Delta \theta$ in the horizontal direction at height $z_p$ and $\langle, \rangle_t$ is a time average. 
The resulting spectra, calculated for $d=100$, $500$, $2250$ and $4096$, are shown in figure~\ref{fig:error_spectra}, alongside the spectrum of the original test dataset as reference. For $d=100$, the error spectrum has large values (compared to the original spectrum) across almost all scales, indicating the inability of the F$d$AE to accurately represent the flow. As the latent space dimension increases, the F$d$AE progressively improves at capturing features corresponding to smaller wavenumbers $k_x$. At $d=2250$, the F$d$AE can faithfully represent scales up to the Kolmogorov scale $k_\eta(z_p)$, therefore completely describing the flow, as expected from the results above. \REFA{The significance of  $z_p$ and the justification for choosing it as the relevant height is given below.}



\begin{figure}
    \centerline{\includegraphics[width=0.85\linewidth]{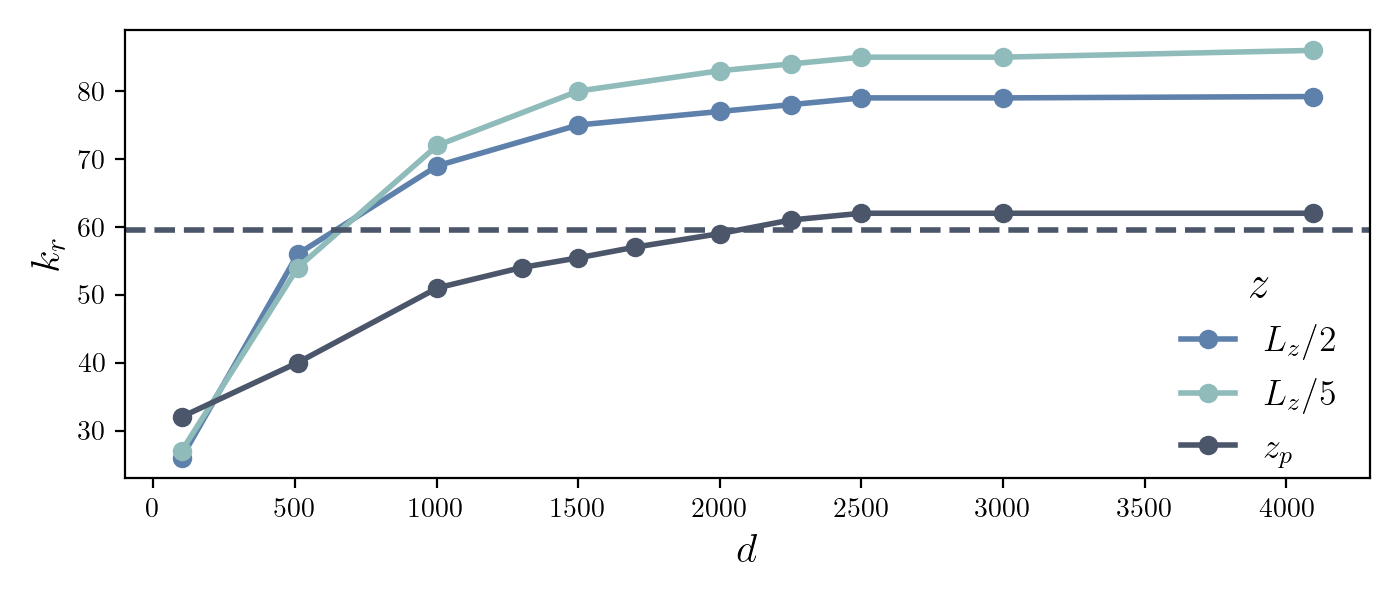}}
    \caption{\REFA{Maximum $k_x^*$ resolved by each F$d$AE for different heights. The horizontal line shows the Kolmogorov scale $k_\eta$ of the flow at $z_p$. At $d^*=2250$ the network reconstructs scales up to $k_\eta$ for $z_p$. }}
    \label{fig:max_kx_diff_spectra}
\end{figure}

By inspecting figure~\ref{fig:error_spectra}, we can define the resolved wavenumber $k_r$ for each $d$ as the first wavenumber $k_x$ at which the corresponding $E_{\Delta \theta}$ intersects with $E_\theta$ (the spectrum of the input). This intersection represents the smallest scale at which the output of the networks fails to accurately reproduce the input, effectively indicating the maximum $k_x$ that is resolved for each $d$. This is shown in figure~\ref{fig:max_kx_diff_spectra}, where we observe that the F$d$AE networks start to reproduce scales up to $k_\eta(z_p)$ for $d=2250$. \REFA{Note that in this figure we also repeat the analysis at heights closer to the center of the channel.
At these higher heights not only is Kolmogorov lengthscale larger, as seen in figure~\ref{fig:eps_eta_theta}, but the autoencoder is also able to resolve smaller scales at lower $d$. We recall too that at lower heights the dynamics are dominated by the boundary layer which is also resolved at lower values of $d$, as shown in figure~\ref{fig:theta_rms}. Therefore, when the autoencoder is able to resolve $\eta_K(z_p)$ it is also able to resolve the smallest scale at every height.}
We then define the minimum dimension of the latent space \REFA{needed by the autoencoder to properly resolve all scales of the temperature field}, $d^*$, as that for which $k_r \approx k_\eta$ \REFA{at $z_p$}. This definition acknowledges the multiscale nature of the flow and does not rely on finding a sharp drop-off in a given metric, a task that is not always possible. \REFA{Note that if the autoencoder was handling other physical fields besides the temperature, this criteria would need to be expanded such that every scale of every field is accounted for.}
Therefore, at $\Ra=10^8$ we obtain $d^*=2250$.  We extend this analysis to other Rayleigh numbers below.


\subsection{Sparsity-inducing Autoencoders}

We now turn our attention to the Sparsity-inducing Autoencoders, SIAE, while focusing again on the $\Ra = 10^8$ case. We repeat the same experiments as above using a fixed bottleneck size of $d'=4096$ and then proceed to analyze the structure of the latent vector. To identify which dimensions of $Z'$ are relevant, we process each snapshot of the test dataset through the trained network, extract $z'$ for each instance, and then calculate the interquartile range (IQR) of each dimension of the vector as a measure of their variability. A large IQR indicates significant activity in that dimension during the reconstruction of the test dataset. 

\begin{figure}
    \centering
    \includegraphics[width=1\linewidth]{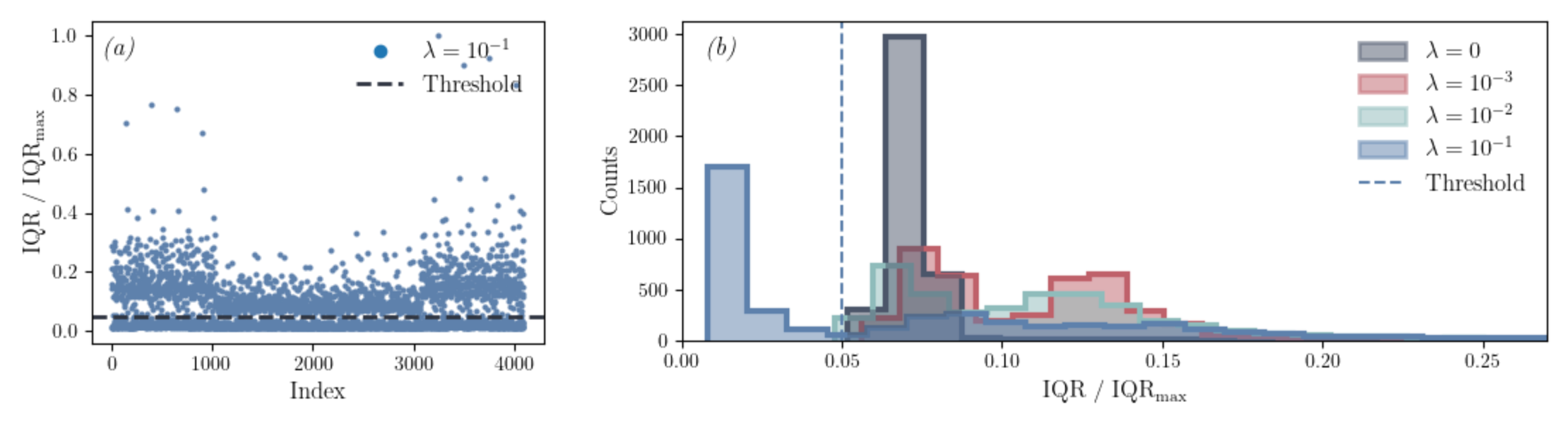}
    \caption{Distribution of the normalized Interquartile Range (IQR) for the SIAE with different regularization values $\lambda$, where the normalization is done using the maximum IQR value. \textit{(a)} IQR values for the network with a regularization strength of $\lambda=10^{-1}$, with the line marking a threshold that differentiates the two observed distributions. \textit{(b)} IQR distributions for various $\lambda$ values, including a comparison with the F$d$AE ($\lambda=0$, shown in darker shading). The line in this figure represents the same threshold as identified in \textit{(a)}. }
    \label{fig:both_histograms_L1}
\end{figure}

We tested different values of the regularization parameter $\lambda$. This led to varying distributions of the interquartile range (IQR) of the bottleneck layer activations as seen in figure~\ref{fig:both_histograms_L1}(b). We employed two criteria for selecting the regularization parameter. First, we evaluated the MSE of the reconstructions for different values of $\lambda$. After this, we analyzed the IQR distributions produced by regularizations that had comparable MSE performance. We considered a good value for $\lambda$ to be one whose bimodal histogram had a left distribution close to zero. This indicates that the neurons in the bottleneck layer corresponding to that distribution were not highly activated. As $\lambda$ is increased, thereby enhancing the regularization strength, the separation between the two groups becomes more pronounced and the left tail of the distribution gets closer to zero IQR, at the expense of reconstruction accuracy as indicated by the MSE.  

Figure~\ref{fig:both_histograms_L1}(a) displays the normalized IQR (normalized by the largest IQR, corresponding to the most active dimension) across each latent space vector index using the chosen regularization strength $\lambda=10^{-1}$. While models such as the Kuramoto-Sivashinsky or damped Korteweg-de Vries, which are characterized by a finite-dimensional manifold, exhibit a distinct separation in the IQR distribution~\citep{sondak_learning_2021}, our high-dimensional Rayleigh-Bénard system does not demonstrate such a clear demarcation at plain sight. Nevertheless, the IQR distribution histograms shown in figure~\ref{fig:both_histograms_L1}(b) indicate two distinct groups of IQR values.

Regardless of the $\lambda$ value for this range, the number of dimensions classified as more relevant (the number of indices whose normalized IQR falls into the second distribution on the right side of the histogram) is around $2000$, these identified dimensions are capable of reconstructing the flow with an MSE comparable to that achieved when using the entire bottleneck layer for decoding. 

\begin{figure}
    \centering
    \includegraphics[width=0.9\linewidth]{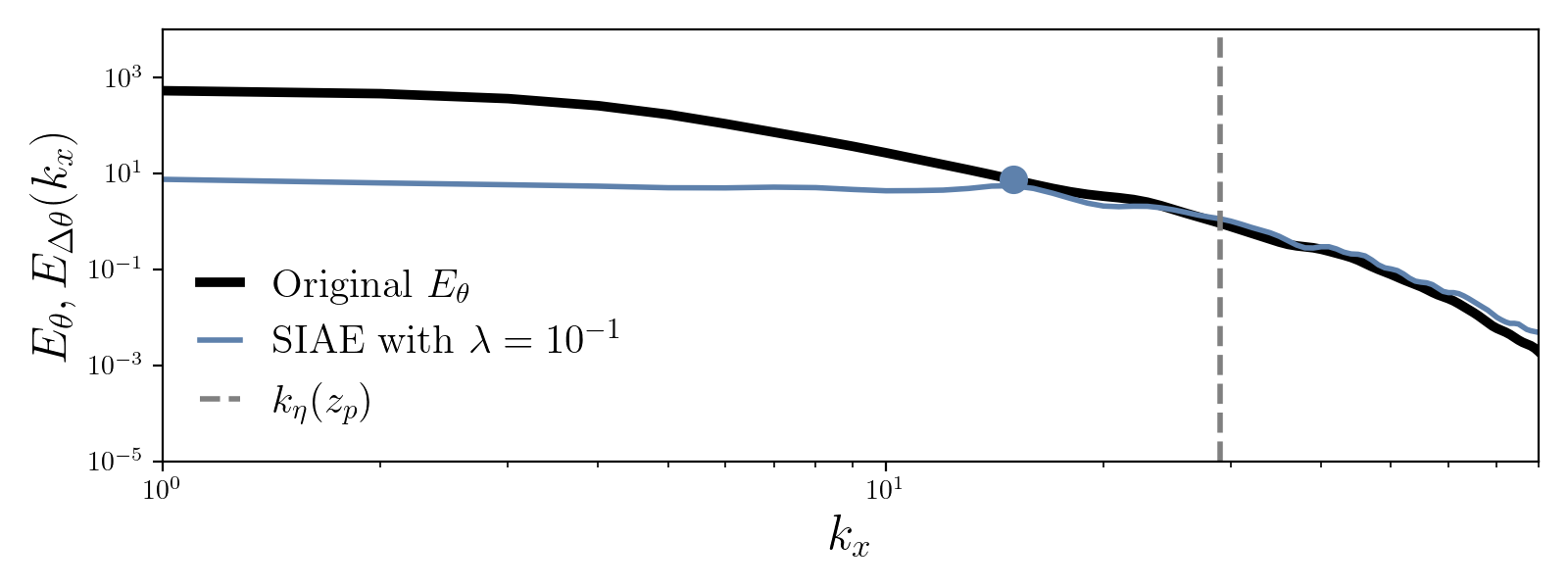}
    \caption{Error spectra $E_{\Delta \theta}$ for SIAE with $\lambda=10^{-1}$, along with the reference energy spectra $E_\theta$ for comparison. The spectra were calculated at $z_p$ by taking the values from the latent space that corresponded to the right distribution of the histogram shown at figure~\ref{fig:both_histograms_L1}(b) \REFA{which has 2010 dimensions.} The dotted vertical line indicates the wavenumber corresponding to the Kolmogorov scale $k_\eta$ at $z_p$, and the dot represents the maximum resolved $k_x$. }
    \label{fig:SIAE_error_spectra}
\end{figure}

SIAE does not provide an ordered set of relevant dimensions in the latent space. Instead, we selected all the values corresponding to the appropriate distribution of IQR (representing the significantly activated latent space dimensions). We then applied the same procedure used for estimating $k_r$ and $d^*$ as described earlier for F$d$AE, as shown in figure~\ref{fig:SIAE_error_spectra} for the case with $\lambda=10^{-1}$. Unlike F$d$AE, SIAE does not reconstruct scales up to $k_\eta$ for every height $z$. This was verified even for $\lambda$ values that performed worse in terms of MSE. For small $\lambda$ values, the criteria is met, but the IQR distribution does not differ significantly from the one observed with $\lambda=0$, resulting in a $d^*$ estimate equal to the total size of the latent space, with no regularization.

\subsection{Implicit regularization}
\label{sec:implicit_regularization}

\begin{figure}
    \centering
    \includegraphics[width=0.85\linewidth]{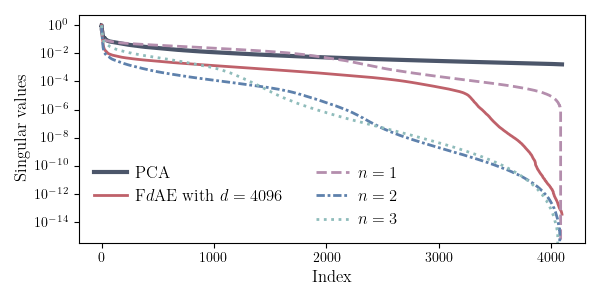}
    \caption{Singular values of the latent space. Each curve represents the singular values of the covariance matrix of the test dataset. Additionally, the thick black line illustrates the spectrum of singular values obtained by performing PCA on the training dataset and projecting the test dataset onto its modes. F$d$AE with $d=4096$ is also shown for comparison. As we increase the number of linear layers $n$, we observe increased regularization.  }
    \label{fig:singular_values_n}
\end{figure}

Finally, we analyze the performance of the IRMAE. Figure~\ref{fig:singular_values_n} shows the singular value spectra of the latent space covariances of $Z'$ for varying numbers of intermediate linear layers $n$. For comparison, we present the singular values obtained from PCA decomposition applied to the training data and then projected onto the test data. Additionally, we examine the spectra from the F$d$AE with $d=4096$, indicative of no implicit regularization, to highlight the distinctions between these models in terms of regularization effectiveness. The PCA singular values remain relatively constant beyond the initial 10 values. However, with an increase in $n$, there is a notable drop in the singular values of IRMAE at certain indices. Specifically, the F$d$AE exhibits a decrease towards the end of the index range, implying the utilization of over 3000 dimensions in the bottleneck layer to represent the data. As $n$ increases, the drop starts to occur at lower index values; with $n=1$, the reduction in singular values is minimal and gradual. In contrast, for $n=2$, the drop is more pronounced. At $n=3$, the regularization becomes too intense, as evidenced by the utilization of $1000$ dimensions, which fails to achieve a effective reconstruction (in terms of the MSE and visual quality) as observed with $n=2$ or lower. For $n=2$, the MSE is $3.5 \times 10^{-5}$, while for $n=3$, it is an order of magnitude larger. When training networks whose dimensionality of the latent vector $z'$ was below the cut-off value, the singular value spectra did not show significant reductions, thereby utilizing the full capacity of the layer.

\begin{figure}
    \centering
    \includegraphics[width=0.9\linewidth]{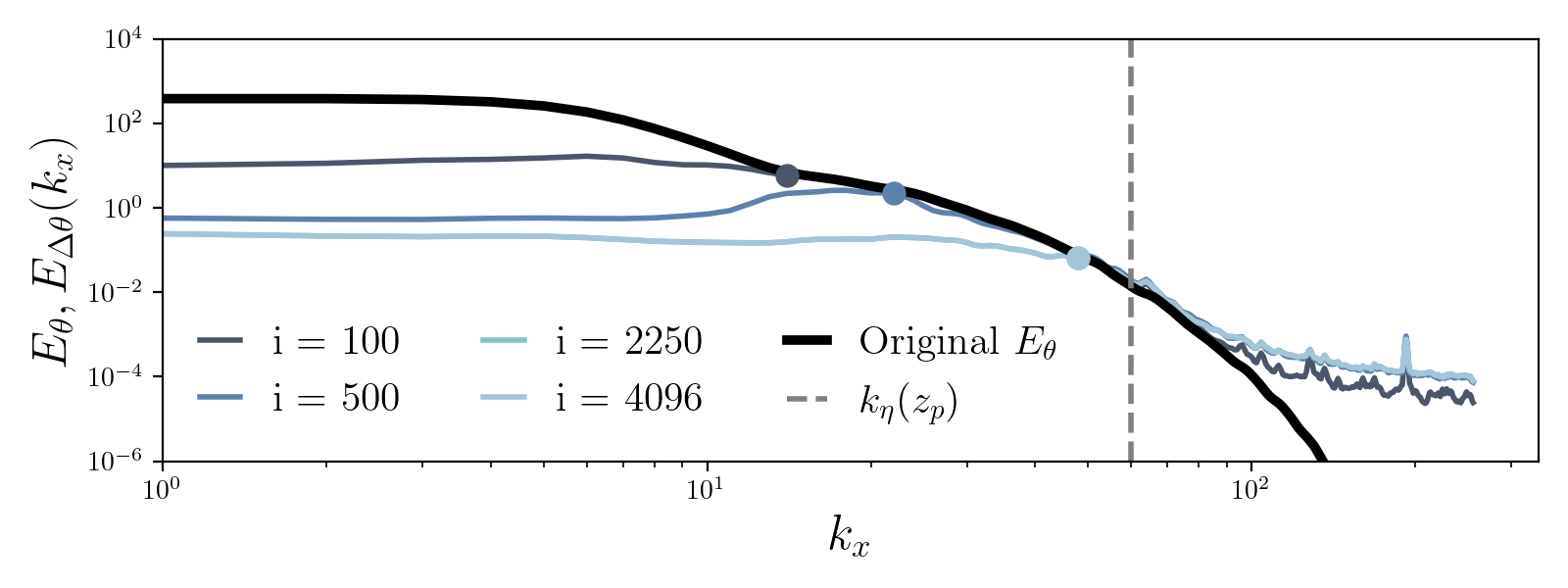}
    \caption{Error spectra $E_{\Delta \theta}$ for IRMAE with $n=2$, along with the reference energy spectra $E_\theta$ for comparison. The spectra were calculated at $z_p$ by taking the first $i$ values from the latent space, based on the singular value spectra. \REFA{Note that the lines for $d=2250$ 
    and $d=4096$ overlap due to the similarity in their spectra, indicating that the IRMAE network does not improve significantly after some $d$.} The dotted vertical line indicates the wavenumber corresponding to the Kolmogorov scale $k_\eta$ at $z_p$, and the dots represent the maximum resolved $k_x$ for each $i$.}
    \label{fig:IRMAE_error_spectra}
\end{figure}

As no sharp drop-off can be observed in the singular value spectra, as was observed in other, smaller, systems \citep{jing_implicit_2020,zeng_autoencoders_2023}, this criterion cannot be used estimate the minimum needed dimension of the latent space. Contrary to SIAE, IRMAE offers a natural ordering of the latent dimensions, and thus the same procedure to estimate $k_r$ and $d^*$ used above for F$d$AE can be repeated here, as seen in figure~\ref{fig:IRMAE_error_spectra} for the case with $n=2$. Doing so
shows that, under this setup (which we tested thoroughly but does not exclude every possible architecture and hyperparameter choice), IRMAE fails to accurately represent all scales, and thus, no estimate of $d^*$ can be extracted. \REFA{We examined the dependence of IRMAE on different values of $N_{\text{train}}$ in Appendix~\ref{appB:IRMAE_N_train} to verify that the inability to reproduce the Kolmogorov scale was not due to the size of the training dataset.}
It is important to note that we were able to get estimates of $d^*$ at lower Rayleigh numbers, but they were consistently larger than the ones obtained via F$d$AE, more details on this below. \REFA{Additionally, the MSE achieved by IRMAE was slightly higher that of F$d$AE, at $5.55 \times 10^{-5}$ and $4.15 \times 10^{-5}$ for $\Ra=10^8$, respectively.}


\subsection{Minimal dimension at different Rayleigh numbers}

Finally, we are interested in estimating the minimal dimension needed to accurately represent every scale \REFA{of the temperature field}, $d^*$, at different Rayleigh numbers by extending he analysis for F$d$AE presented in Section~\ref{subsec:FdAE}. This can be done without any modifications to the method outlined. As examples, the cases with $Ra=10^6$ and $10^7$ are shown in Appendix~\ref{appA:Low_Ra}. 
Figure~\ref{fig:d_vs_ra} shows $d^*$ as a function of the Rayleigh number. The uncertainty in the F$d$AE estimates is due to the discretization of the latent dimension sizes tested. For instance, for Ra=$10^7$, the estimate for $d^*$ was 75, based on tests with $d=50$, $d=75$, and $d=100$.
Since $d=75$ was the first dimension to meet the criteria, the uncertainty is taken as 25, representing the interval between the tested dimensions.
This error likely overestimates the true uncertainty. 
We also show $n_A$, an estimate of the degrees of freedom of the \REFA{field} calculated by taking the area of the domain and dividing it by the smallest physical area $\delta_\theta \, \eta_K$.
As we could not get IRMAE to work for the whole range of Rayleigh numbers studied, and as the values we were able to obtain were always considerably larger than those obtained via F$d$AE (for example, we obtained values of 50 and 450 for $\Ra=10^6$ and $10^7$, respectively), we decided to not include them in the figure.
The results obtained via F$d$AE show a clear transition in the dynamics of the flow between $\Ra=5 \times 10^6$ and $10^7$.
Below the transition $d^*$ has a value of 5, around the transition it starts to grow quickly, and at around $\Ra=2.5 \times 10^7$, it slows down again and $d^*$ appears starts to scale as $n_A$. Unsurprisingly, $n_A$ does not exhibit different behaviours before and after the transition, as the scaling of $\delta_\theta$ and $\eta_K$ does not change \citep{Scheel_2013,  siggia_high_1994, grossmann_scaling_2000}.

\begin{figure}
    \centering
    \includegraphics[width=0.9\linewidth]{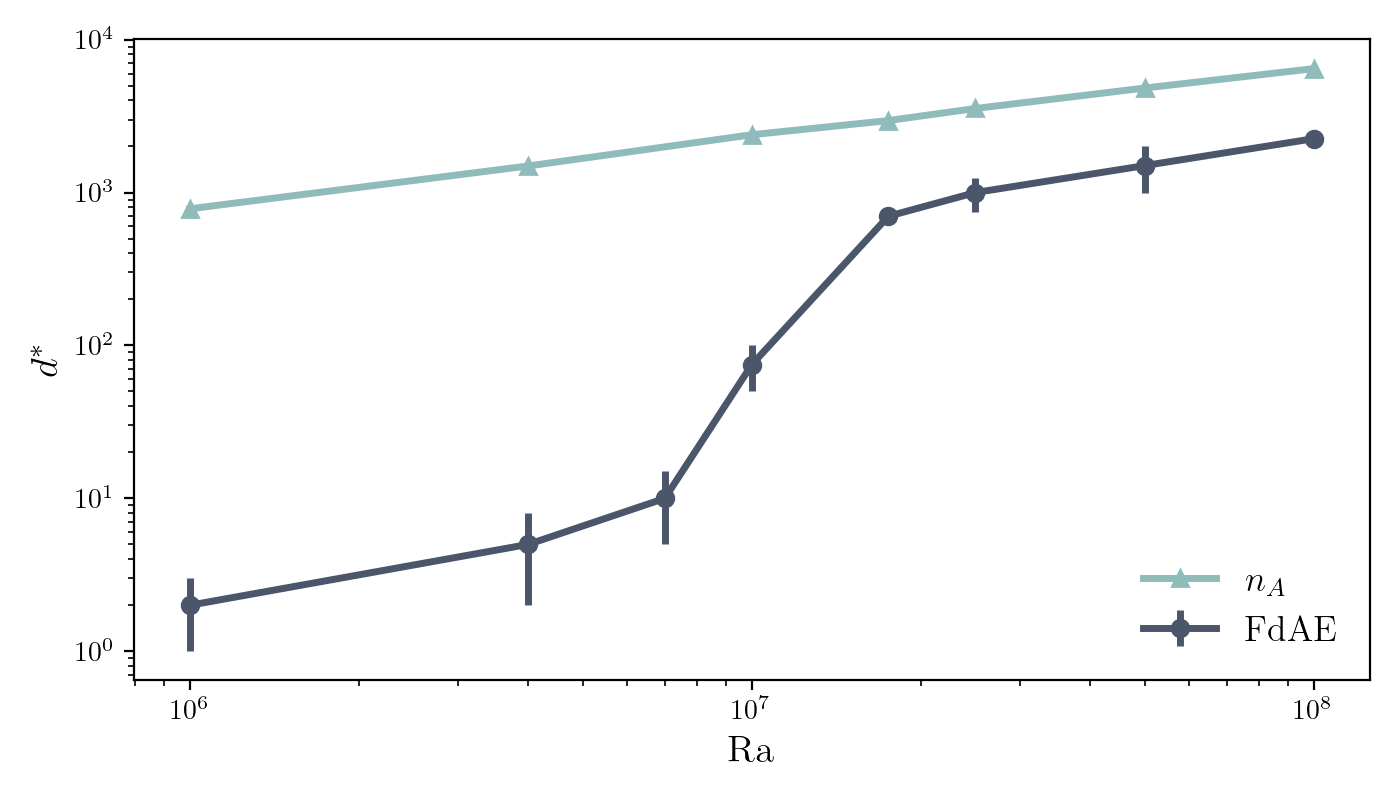}
    \caption{Minimum dimension needed to represent all relevant physical scales \REFA{of the temperature field} using F$d$AE, $d^*$, as a function of the Rayleigh number. An estimate of the number of degrees of freedom that scales as the inverse of $\delta_\theta\;\eta_K$ is also shown.}
    \label{fig:d_vs_ra}
\end{figure}

\section{Conclusions}
\label{sec:conclusions}

In this study we analyzed the performance of different autoencoder architectures when trained to generate reduced-order representations \REFA{of the temperature field} of Rayleigh-Bénard flows \REFA{as the flow transitions to turbulence}.
We quantified the error at each scale, and defined $d^*$ as that needed to obtain a small error at every physically-relevant scale.
This criterion ultimately worked much better for multiscale flows than previously suggested ones.
We used this methodology to calculate $d^*$ at Rayleigh numbers ranging from $10^6$ to $10^8$.
The results indicate a sharp increase in dimensionality as the flow transitions to turbulence, with $d^*$ being equal to 2 at $\Ra=10^6$, and rapidly climbing several orders of magnitude after, reaching a value of 2250 at $\Ra=10^8$.
\REFA{Note that in this work we focused only on the temperature field. The values of $d^*$ obtained serve then only as a proxy for the dimensionality of the whole system but are still indicative of its behaviour. Following past results in the synchronization properties of Rayleigh-Bénard flows \citep{agasthya_reconstructing_2022}, it can be expected that at low $\Ra$ the value of $d^*$ obtained here for the temperature field is the same as that of the whole system, as the velocity and temperature fields are locked, while at higher $\Ra$ the coupling between the two fields decreases and dimensionality of the whole system is probably larger than just that of the temperature field. These statements will be checked empirically in future work.}
Having a parametrization of the submanifold where the dynamics live that captures the transition to turbulence opens an important door in studying how the mechanisms of the transitions that will be explored in future work.

We also tested autoencoder architectures developed for fluid dynamics dimensionality reduction by \cite{sondak_learning_2021} and \cite{zeng_autoencoders_2023}, the SIAE and the IRMAE, respectively. Both architectures incorporate regularizations in the latent space to minimize its effective dimensionality. Despite their success with simpler systems, they failed to produce satisfactory results at every Rayleigh number studied here. At the Rayleigh numbers where they did work, they produced consistently higher results than F$d$AE. The quest for regularized architectures that can estimate $d^*$ accurately in highly turbulent and multiscale flows remains open.

The dimensionality reduction achieved here gives an opportunity to employ Neural Ordinary Differential Equations (Neural ODEs) for dynamic modeling. Neural ODEs could provide a powerful tool for modeling the time-evolution of the flow's latent features, without needing the full computational complexity of the higher-dimensional space.

The authors would like to thank Gabriel Torre, Pablo Mininni, Michele Buzzicotti, Luca Biferale and Mauro Sbragaglia for fruitful discussions. This work was partially supported by the Google PhD Fellowship Program.

Code is available at \url{https://github.com/melivinograd/FdAE\_SIAE\_IRMAE}.

\appendix
\section{Results for Lower Rayleigh Numbers ($\Ra=10^6$ and $10^7$)}\label{appA:Low_Ra}

In the main text, we focused on presenting detailed results for the highest and most complex Rayleigh number, $\Ra = 10^8$, among the values considered. To provide a broader perspective on the system's behavior at different Rayleigh numbers, this appendix presents additional results for $\Ra = 10^6$ and $\Ra = 10^7$, where the minimum dimensions needed to represent the smallest scales of fluid dynamics are $d^* = 2$ and $d^* = 75$, respectively.

\begin{figure}
    \centering
    \includegraphics[width=0.9\linewidth]{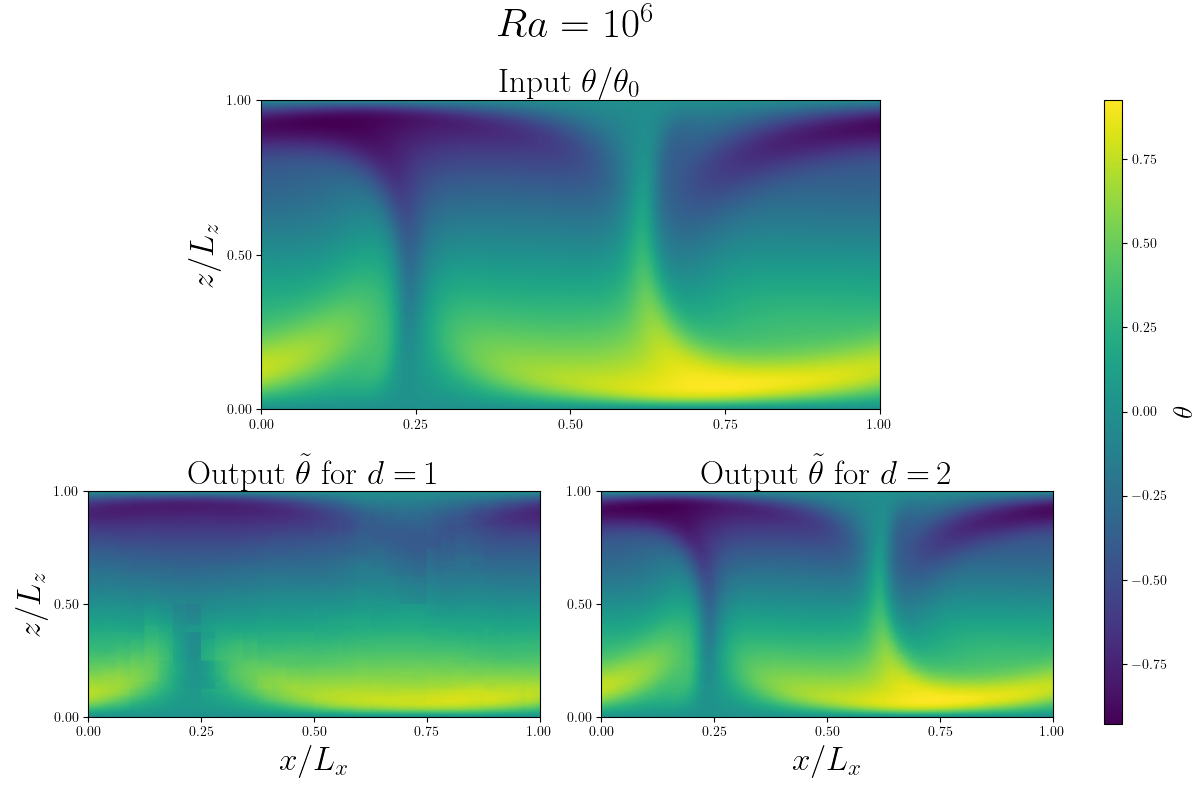}
\caption{Output fields $\tilde{\theta}$ of the FdAE for $\Ra=10^6$, showing a snapshot of the output temperature field for latent space dimension $d = \{1, 2\}$. The value $d=2$ corresponds to $d^*$ for this $\Ra$. The normalized ground truth field, $\theta / \theta_0$, used as input, is shown for comparison.}
    \label{fig:1e6_visualization}
\end{figure}

\begin{figure}
    \centerline{\includegraphics[width=0.9\linewidth]{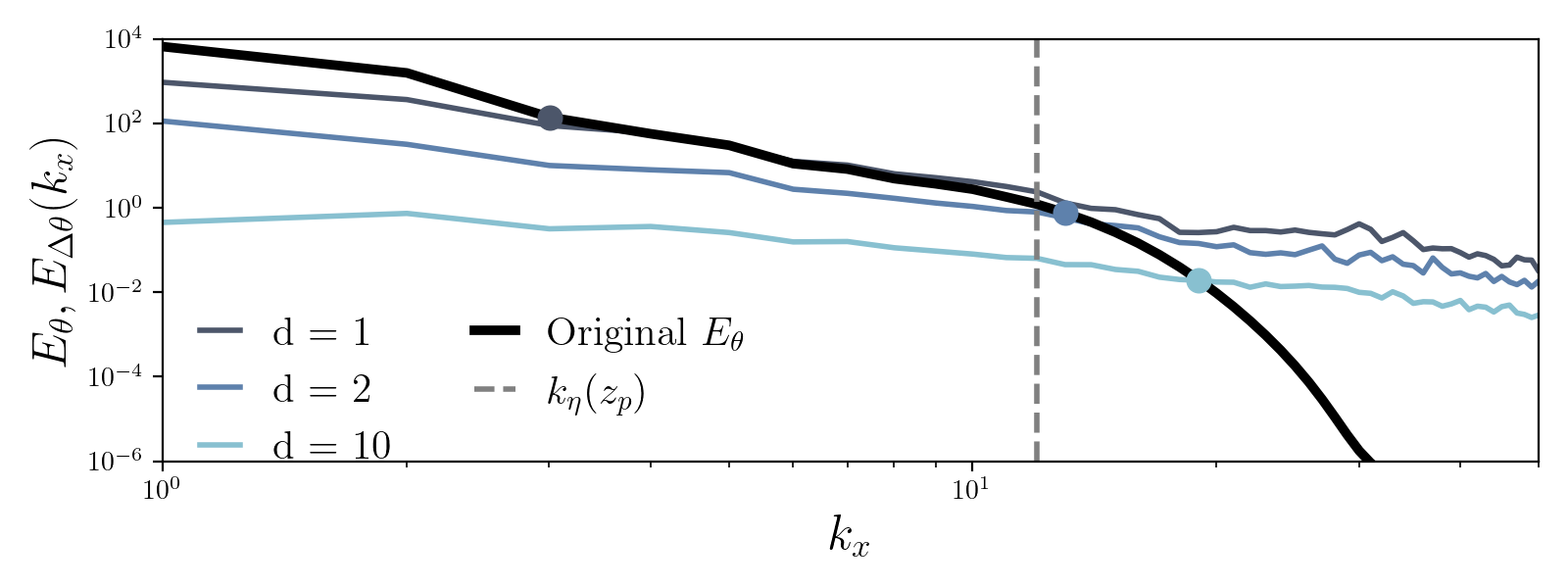}}
    \caption{Error spectra $E_{\Delta \theta}$ for various reconstructions, alongside the reference energy spectra $E_\theta$ for comparison. The spectra were calculated at $z_p$. The dotted vertical line indicates the wavenumber corresponding to the Kolmogorov scale $k_\eta$ also at $z_p$, and the dots represent the maximum resolved $k_x$ for each $d$.}
    \label{fig:1e6_energy_difference}
\end{figure}

\begin{figure}
    \centering
    \includegraphics[width=0.9\linewidth]{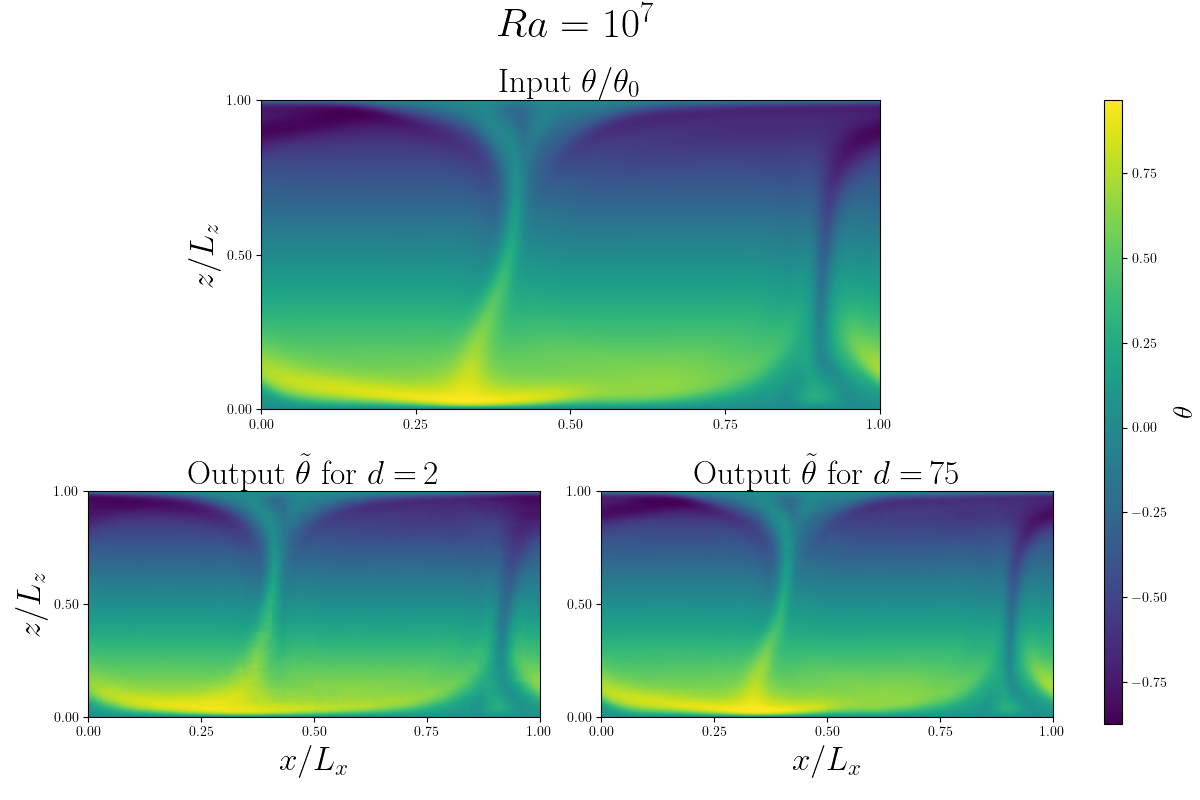}
\caption{Output fields $\tilde{\theta}$ of the FdAE for $\Ra=10^7$, showing a snapshot of the output temperature field for latent space dimensions $d = \{10, 75\}$. The value $d=75$ corresponds to $d^*$ for this $\Ra$. The normalized ground truth field, $\theta / \theta_0$, used as input, is shown for comparison.}

    \label{fig:1e7_visualization}
\end{figure}

\begin{figure}
    \centerline{\includegraphics[width=0.9\linewidth]{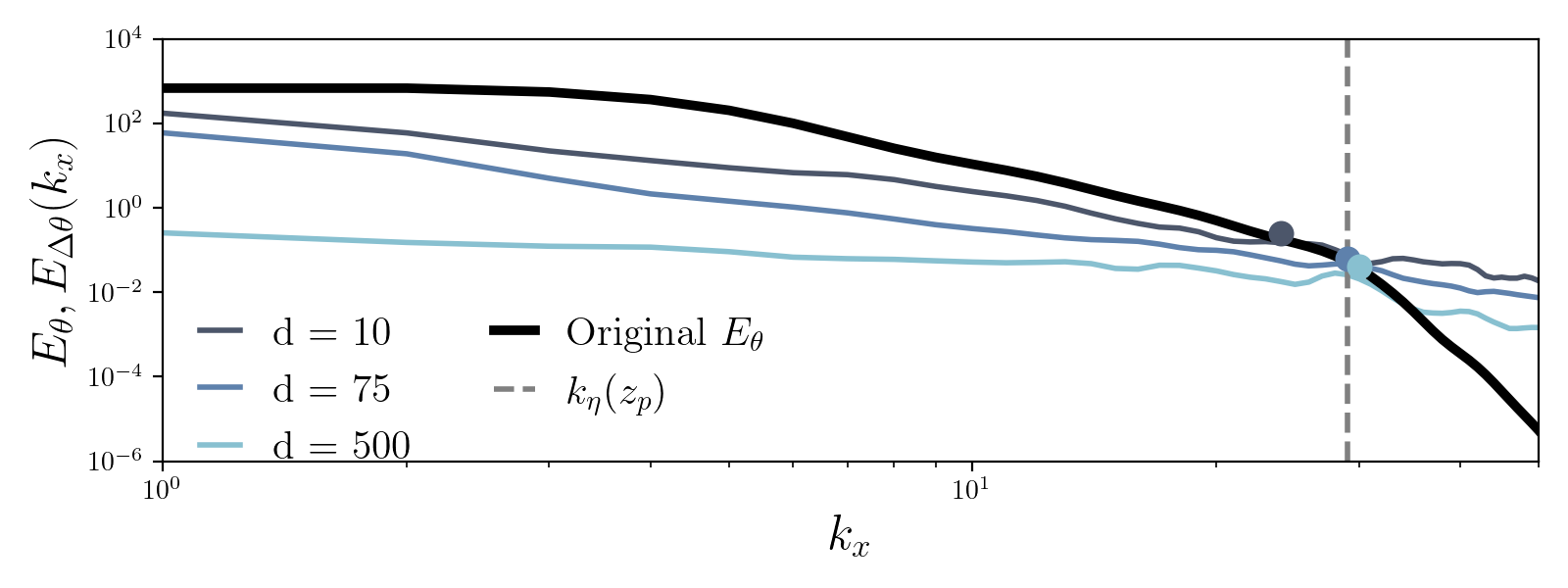}}
    \caption{Error spectra $E_{\Delta \theta}$ for various reconstructions, alongside the reference energy spectra $E_\theta$ for comparison. The spectra were calculated at $z_p$. The dotted vertical line indicates the wavenumber corresponding to the Kolmogorov scale $k_\eta$ also at $z_p$, and the dots represent the maximum resolved $k_x$ for each $d$.}
    \label{fig:1e7_energy_difference}
\end{figure}

\begin{figure}
    \centerline{\includegraphics[width=0.85\linewidth]{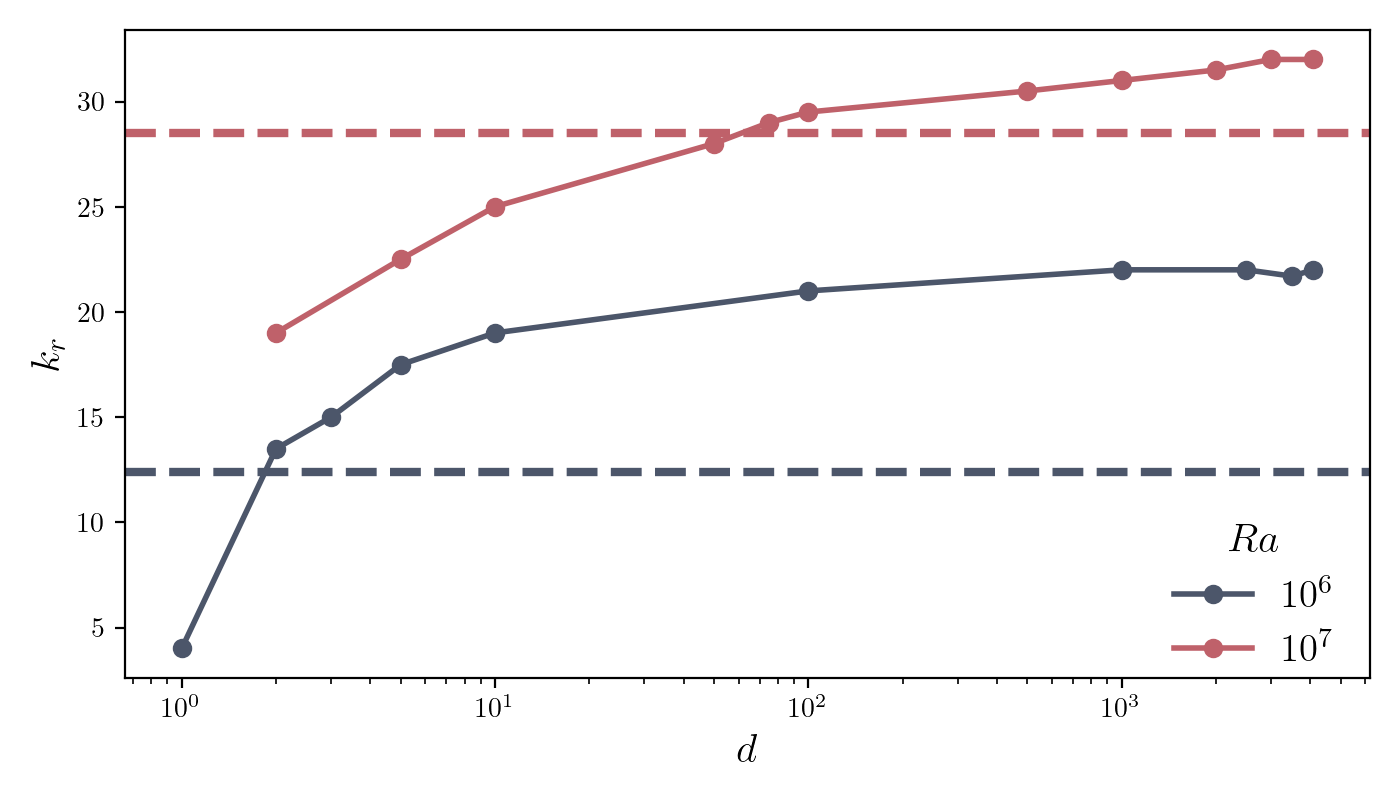}}
    \caption{\REFB{Maximum $k_x^*$ resolved by each F$d$AE for $\Ra=10^6$ and $\Ra=10^7$, as seen in figure~\ref{fig:max_kx_diff_spectra}. The horizontal dashed lines show the Kolmogorov scale $k_\eta$ of the flow at each $z_p$ for both Rayleigh numbers. }}
    \label{fig:kr_1e6_1e7}
\end{figure}

Here, we show visualizations of the F$d$AE output at different latent space dimensions, $d$, as seen in figure~\ref{fig:visualizations}. The original normalized temperature field used as input for the networks is also presented for reference. For each $\Ra$, we display two values of $d$: one below the corresponding $d^*$ and the other at $d = d^*$, as shown in figures~\ref{fig:1e6_visualization} and~\ref{fig:1e7_visualization} for $\Ra = 10^6$ and $\Ra = 10^7$.
We also present analogous figures to figure~\ref{fig:error_spectra} for these two $\Ra$ values. For each case, we show the original energy spectrum $E_\theta$ and the difference spectra $E_{\Delta\theta}$ for different latent space dimensions, one at $d = d^*$, and one higher than $d^*$ as shown in figures~\ref{fig:1e6_energy_difference} and~\ref{fig:1e7_energy_difference}.
\REFB{
Finally, we include the analogous to figure~\ref{fig:max_kx_diff_spectra} or $\Ra=10^6$ and $\Ra=10^7$ in figure~\ref{fig:kr_1e6_1e7}, which illustrates the maximum resolved wavenumber $k_r$ achieved by each F$d$AE model across different latent space dimensions. 
}

\section{Impact of Training Dataset Size on IRMAE Performance}\label{appB:IRMAE_N_train}
\REFA{
At the highest Rayleigh number considered, $\Ra = 10^8$, IRMAE failed to accurately represent all scales of the flow, and thus, no estimate of $d^*$ can be extracted. We conducted several experiments at the highest Rayleigh number considered, $\Ra = 10^8$, to verify this. In these experiments, we varied the amount of data used for training, denoted by $N_{\text{train}}$. The performance of the trained autoencoders was evaluated using a consistent test set, matching the size of the test set used throughout the study. This procedure was applied to IRMAE autoencoders, leveraging the fact that we can train a single network for all $d$ values.}

\begin{figure}
    \centering
    \includegraphics[width=\linewidth]{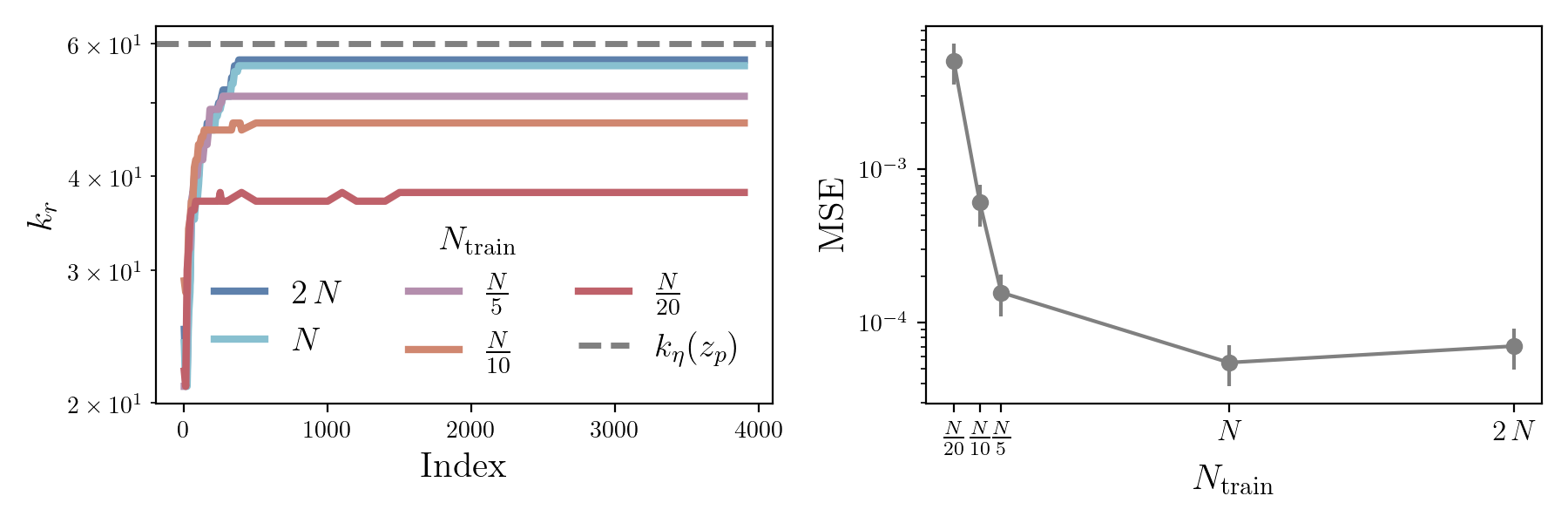}
    \caption{\REFA{(a) Maximum wavenumber resolved by each IRMAE network for different numbers of training images, $N_{\text{train}}$, at $\Ra=10^8$. The horizontal line indicates the Kolmogorov scale $k_\eta$ for this Rayleigh number. (b) Mean Squared Error (MSE) of the validation dataset for the temperature field for each network.}}
    \label{fig:IRMAE_N_train}
\end{figure}

\REFA{
As shown in figure~\ref{fig:IRMAE_N_train}, the performance of the autoencoders, measured in terms of the maximum resolved wavenumber $k_r$, improves with an increase in $N_{train}$. It is important to note that all the autoencoders exhibit similar performance for small wavenumbers $k$, except for the case with a low amount of training data ($N_{train} = 100$, augmented with 10 rolls). Indeed, increasing $N_{train}$ results in the reconstruction of more $k_x$ values, but even with a dataset twice as large, we were unable to resolve scales up to $k_\eta$, with only marginal improvement in the maximum resolved $k_x$. We also evaluated the mean squared error (MSE) across these experiments, observing minimal improvement as $N_{\text{train}}$ increases.
}

\bibliographystyle{jfm}
\bibliography{bib}

\end{document}